\documentclass[aps,prd,a4paper,nosuperscriptaddress,nofootinbib,showpacs,twocolumn,showkeys,amsfonts,amssymb,amsmath]{revtex4-2}
\usepackage{amssymb,latexsym}

\usepackage{color}
\usepackage{amscd}
\usepackage{times}
\usepackage{epsfig}
\usepackage{psfrag}
\usepackage{graphicx}
\usepackage{amsthm,amsmath}
\usepackage{mathrsfs}
\usepackage[colorlinks=true, allcolors=blue]{hyperref}
\usepackage{leftindex}

\begin{document}
\title[]{Gyroscopic stability for nanoparticles in Stern-Gerlach Interferometry and spin contrast
}
\newcommand{\affone}{University College London, Gower Street, WC1E 6BT London, United Kingdom}
\newcommand{\affthree}{Van Swinderen Institute, University of Groningen, 9747 AG, The Netherlands}

\author{Tian Zhou $^{1}$}
\author{Sougato Bose $^{2}$}
\author{Anupam Mazumdar $^{1}$}
\affiliation{ 
$^{1}$ Van Swinderen Institute, University of Groningen, 9747 AG Groningen, The Netherlands\\
 $^{2}$~University College London, Gower Street, WC1E 6BT London, United Kingdom}

\begin{abstract}

There is a severe impediment in Stern-Gerlach interferometry with a spin-embedded nanoparticle. This is because particle rotation creates a significant mismatch of the wave function of angular degrees of freedom during the creation and recombination of the spatial superposition, resulting in an inferior visibility contrast. We show that imparting angular momentum along the direction of a defect, such as the nitrogen-vacancy (NV) center, during the Stern-Gerlach interferometry provides excellent gyroscopic quantum stability to a nanodiamond, which opens up a way to maintain significant visibility in the quantum interference.  The spin contrast improves by many folds for a wide-ranging value of the initial angular momentum, e.g. $10^{3}-10^{6}$~Hz for a mass of order $10^{-14}-10^{-17}$ kg nanodiamond. Our scheme is independent of the material and is pertinent to any nanoparticle that can be used to create massive and large spatial superpositions with an embedded spin defect.

\end{abstract}

\maketitle

\section{Introduction}

Stern-Gerlach interferometry(SGI)~\cite{SGI_experiment,amit2019t,WanPRA16_GM,Scala13_GM,Pedernales:2020nmf,Marshman:2021wyk} with nanoparticles, in particular in the context of a nitrogen-vacancy (NV) center of nanodiamonds (ND), is a method of creating the spatial superposition and have ubiquitous applications, including quantum sensors~\cite{Wu:2022rdv}, detection of dark matter particles~\cite{Barker:2022mdz}, gravitational waves~\cite{Marshman:2018upe}, testing the quantum nature of gravity in a lab~\cite{Bose:2017nin,marshman2020locality,bose2023entanglement,Carney_2019,Carney:2021vvt,Danielson:2021egj,christodoulou2023locally,Schut:2021svd,Tilly:2021qef,Marletto:2017kzi}, short-distance Casimir and dipole entanglement mediated via photon~\cite{vandeKamp:2020rqh,marshman2024entanglement}, extra dimensions~\cite{elahi2023probing}, equivalence principle~\cite{bose2023entanglement,chakraborty2023distinguishing}, post-Newtonian gravity~\cite{Toros:2024ozf}, relativistic effects in quantum electrodynamics~\cite{Toros:2024ozu}, and tests beyond general relativity~\cite{Vinckers:2023grv}. However, there remains a demanding requirement on quantum spatial superpositions of distinct localized states of neutral mesoscopic masses $m \sim 10^{-14} - 10^{-17}$ kg over spatial separations of $\Delta x \sim {\cal O}(10-50) {\rm \mu m}$~\cite{Bose:2017nin,Schut:2023hsy,Schut:2023eux}, far beyond the scales achieved to date (e.g., macromolecules $m \sim  10^{-22}$ kg over $\Delta x \sim 0.25 {\rm \mu m}$, or atoms $m\sim 10^{-25}$ kg over $\Delta x\sim 0.5$ m~\cite{arndt,overstreet2022observation,asenbaum2017phase}. 

 

To finish the final spin measurement of SGI, one of the critical challenges is maintaining the coherence (or contrast) of the NV spin. There are many sources of coherence loss. The most studied coherence loss is the decoherence effect arising from the interaction between the ND system and the environment, such as collisions by air molecules, electromagnetic noise and spin-spin interaction between the NV spin and another spin in the ND or the environment~\cite{Bose:2017nin,vandeKamp:2020rqh,Rijavec:2020qxd,Schut:2021svd,Schut:2023hsy,Schut:2023eux,Fragolino:2023agd,Schut:2023tce,Zhou:2022epb,Zhou:2022frl,Zhou:2022jug}. These decoherence effects can be mitigated by optimizing the experimental environment, such as applying higher vacuum, lower temperature, and better interaction screens for the SGI system. However, even if the SGI is working in an ideal environment, there is still coherence loss of the NV spin, known as the 
Humpty-Dumpty (H-D) effect~\cite{Englert,Schwinger}. It arises from the ND's wave packet mismatch of the two SGI arms.


For the first time, the role of libration mode has been studied recently in the context of the H-D problem, e.g. loss of spin contrast \cite{Japha:2022phw, japha2022role}. The authors pointed out that, because of the NV center's specific Hamiltonian, the spin dynamics manifests the ND in the libration mode due to the angular mismatch between the external magnetic field and the NV axis. The libration mode gives rise to two significant effects; it selects the spin states to avoid an unstable libration mode of the ND and severely affects the spin contrast. In the former case, the authors showed that the stable states are $|0\rangle,|-1\rangle$ (among the $|0\rangle,|+1\rangle,|-1\rangle$ states) for the choice of the libration mode around its minimum, e.g. the libration angle $\theta =0$. This penalizes the size of the spatial superposition by half, which is one of the crucial goals to achieve in fundamental experiments. More remarkably, it causes asymmetric evolutions of the wave packets of the two SGI arms, which is harmful to the spin contrast (also called the "quantum uncertainty limit" in \cite{Japha:2022phw}). To obtain a large spin contrast, in~\cite{Japha:2022phw, japha2022role}, the authors proposed a special SGI scheme with a short duration ($t\sim100\mu s$) and a relatively small spherical object with a mass ($\sim 10^{-19}$~kg) of ND. In this case, they neglected the diamagnetic force because the mass of the ND is small, and they assumed that the force acting on the center of mass(CoM) of the ND and the torque acting on the ND can both be considered constants. However, none of these assumptions is applicable to a more massive object, which causes more limitations in the applications of their SGI design.

Instead of treating it as a problem, we will provide a solution to mitigate spin contrast loss while maintaining the superposition size governed by both states $|+1\rangle,|-1\rangle$ for general SGI models with a heavy nanoparticle of mass such as $10^{-14}-10^{-17}$~kg, where we also include the effect of diamagnetic-induced potential in the Hamiltonian. The critical result of this paper can be summarized as follows: (1) If the nanoparticle is initiated with an initial angular momentum along the direction of the spatial superposition, we would be able to use both $|+1\rangle, |-1\rangle$ spin states, which would favour the size of the superposition. (2) More importantly, an initial angular momentum will aid in obtaining a better spin contrast. (3) Furthermore, having a considerable initial angular momentum can also help us to average any deleterious effect due to permanent dipoles, as shown in \cite{Afek:2021bua}. In the latter case, the authors observed that the dipole effects are mitigated while the nanoparticle rotates. Hence, for the first time, we can minimize the deleterious effects of a spin-centered nanoparticle in creating and observing the spatial superposition in a lab.

In section II, we will discuss a simple model of spatial superposition in one dimension via the Stern-Gerlach protocol~\cite{Marshman:2021wyk}. In section III, we will treat the nanoparticle, e.g. ND as a rigid body, and include the effect of rotation along the direction of the NV-center to study the dynamics of the Euler angles during the process of creation and annihilation of the spatial quantum. In section IV, we will discuss the spin contrast for the full one-loop interferometry, including thermal effects in the libration mode of the ND. Finally, we will conclude in section V. 


\section{A model of spatial superposition}\label{ModelofSGI}

In this section, we will give an illustrative model of spatial superposition via the Stern-Gerlach setup. We will follow the model given by \cite{Marshman:2021wyk}.
We consider a levitated ND with a single spin-1 NV center, which the Hamiltonian is given by~\cite{Pedernales:2020nmf,Marshman:2021wyk,Japha:2022phw}:
\begin{equation}\label{hamiltonian}
    H  = \frac{\textbf{p}^2}{2 m} +\sum_{j=1}^3 \frac{L^2_j}{2I_j} + \frac{-\chi_\rho m}{2 \mu_0}\textbf{B}^2 + \mu\,\textbf{s}\cdot \textbf{B} + D S_{\parallel}^2\, ,
\end{equation}
where $m$ and $I_j$ are the mass and inertia moments of the ND respectively, $\chi_\rho=-6.2\times 10^{-9}m^3/kg$ represent the mass magnetic susceptibility of diamonds, the magnetic moment of the NV is $\mu=e\hbar/2m_e=h\times2.8{\rm MHz/G}$, and $D=h\times2.87{\rm GHz}$ is the zero-field splitting of the NV center, $h=6.62607015\times10^{-34}{\rm JHz^{-1}}$, $S_{\parallel}=\textbf{s}\cdot\hat{n}_s$ is the projection of the NV spin on the NV crystalline direction, denoted as $\hat{n}_s$.


We will follow the SGI scheme proposed in Ref.\cite{Marshman:2021wyk} as an example, where the following time-varying spatial magnetic field is applied:
\begin{equation}\label{magneticfield}
\vec{B}(z,x)=\left\{ 
\begin{aligned}
    &(B_0-\eta z)\hat{z}+\eta x\hat{x}\, ,\qquad t<\tau_1\\
    &\qquad\qquad B_1 \hat{z}\, , \qquad \quad\quad \tau_1\le t< \tau_2 \\
    &-(B_0-\eta z)\hat{z}-\eta x\hat{x}\, , \quad t>\tau_2
    \end{aligned}
\right.
\end{equation}
in which $B_0$, $B_1$, and $\eta$ are constants. In \cite{Marshman:2021wyk}, initially, the ND is set as translationally and rotationally static at the original point $z=0$ and $x=0$. Meanwhile, the NV axis $\hat{n}_s $ is oriented in the $\hat{z}$ direction. We also choose the NV axis as the quantization orientation of the NV spin state, and the spin state is initially prepared as the superposition state $|s\rangle=(\lvert+1\rangle+\lvert-1\rangle)/\sqrt{2}$ of the NV spin. The large splitting $D$ of the energy level prevents the NV spin flipping to the $s=0$ state due to the $\mu\eta x s_x$ term in the Hamiltonian. So, the NV spin state is stable during the experiment.

In \cite{Marshman:2021wyk}, the rotation of ND is not considered, so it is always assumed that the NV axis is oriented to the direction $\hat{z}$. We will relax this condition; see below. In addition, since $x=0$ is the equilibrium position of the harmonic potential in the direction $\hat{x}$, the translational motion along the $\hat{x}$ axis is always static except for initial thermal noise, which can be reduced by cooling the initial oscillation of the ND~\cite{Deli__2020,gieseler2012subkelvin,Piotrowski_2023}. This can be arranged in a narrow trap where the motion is restricted to effectively one dimension\cite{Schut:2023hsy}. Meanwhile, the NV spin state results in a zero expectation value of $s_x$. So, there is no force from the Zeeman term in the $x$ direction. Therefore, the motion along the $\hat{x}$ direction is negligible. 

The homogeneous magnetic field $B_1\hat{z}$ set during $\tau_1$ to $\tau_2$ aims to ensure that the ND is always in enough sizeable magnetic field to avoid Majorana spin flipping~\cite{Marshman:2021wyk}. Based on the initial conditions and assumptions above, we realize a 1-dimensional SGI, where only the motion along the $\hat{z}$ axis should be considered. Then, from the Hamiltonian (\ref{hamiltonian}) and the magnetic field (\ref{magneticfield}), we obtain the effective Hamiltonian for the motion of ND in the $z$ direction:
\begin{equation}\label{H-O}
    H_z=\frac{\hat{p}_z^2}{2m}+\frac{1}{2}m\Omega^2(\hat{z}-Z_0)^2-\tilde{\eta}(t)\mu\hat{s}(\hat{z}-Z_0)\, .
\end{equation}
in which $\Omega$ and $Z_0$ is defined as $\Omega^2=-\chi_\rho \tilde{\eta}^2/\mu_0$, $Z_0=B_0/\eta$, and $\tilde{\eta}(t)$ is the magnetic field gradient, namely $\tilde{\eta}(t)=-\eta,~0,~+\eta$ when $t<\tau_1$, $\tau_1\le t<\tau_2$, and $t\ge\tau_2$, respectively. As we can see, the ND is in a spatial harmonic potential when the linearized magnetic field is applied, and the equilibrium position of the harmonic potential depends on the NV spin state, e.g. the last term $\hat{s}$ of Eq.~\ref{H-O}. The largest spatial size of superposition during the SGI interferometer can be estimated by the distance between the equilibrium position of the $s=\pm 1$ state, namely $\Delta z\approx2\mu\eta/m\Omega^2=2\mu_0\mu/(-\chi_\rho m \eta)$~\cite{Pedernales:2020nmf,Marshman:2021wyk}, which is inversely proportional to the magnetic field gradient $\eta$ and the mass $m$ of ND. The time scale of the interferometer is about $\Omega\propto\eta^{-1}$. 
To close the two SGI arms, a rapid microwave pulse should flip the NV spin state at an appropriate value of $\tau_3$. Afterwards, the interferometer is closed at a specific time, denoted $\tau_4$. In our specific example, we take $\tau_4\sim 1$s, which this experimental time scale is also assumed in many SGI protocols \cite{Bose:2017nin,Marshman:2021wyk,Pedernales:2020nmf}. Moreover, this time can be made smaller, albeit at the cost of a smaller spatial superposition. It depends on the coherence time of the embedded spin in a nanoparticle. The coherence time is increasingly getting better for NV centered systems. For instance, the NV spin can be mapped to the nuclear spin of $^{13}C$ atoms to get a better coherence time, see~\cite{Abobeih_2018}.


Remarkably, since the $s=+1$ state is unstable, as shown in Refs.~\cite{Japha:2022phw,japha2022role}, the ND with $s=+1$ spin state will have a significant rotation, so the NV spin is no longer oriented to the $\hat{z}$ axis, Then the spatial trajectory will be modified a lot. 
To address this problem, a possible way is to create a spatial superposition by $|s\rangle=(\lvert0\rangle+\lvert-1\rangle)/\sqrt{2}$, because the ND with the $s=0$ state is torque-free and with the $s=-1$ state is rotationally stable~\cite{Japha:2022phw,japha2022role}. This novel approach ensures that the NV spin is always approximately oriented to the $\hat{z}$ axis; however, the price is that the superposition size is halved and, to make matters worse, the spin contrast is almost zero because of the mismatch between the two SGI angular paths and the mismatch of the quantum uncertainty of the two asymmetric angular wave packets(see \cite{Japha:2022phw} and Appendix \ref{appA}). To reduce the contrast loss, schemes using the $|s\rangle=(\lvert0\rangle+\lvert-1\rangle)/\sqrt{2}$ state require extreme constraints for the location of the NV spin in the ND, e.g. within nanometer separation distance of the NV center from the COM of the ND, and very difficult fine-tuning of experimental parameters, see~\cite{Japha:2022phw}. For completeness, we illustrate this situation for our SGI model, Eq.~\ref{magneticfield}, see Ref.~\cite{Marshman:2021wyk}, in Appendix \ref{appA}.

\begin{figure}       
\centering
\includegraphics[width=\linewidth]{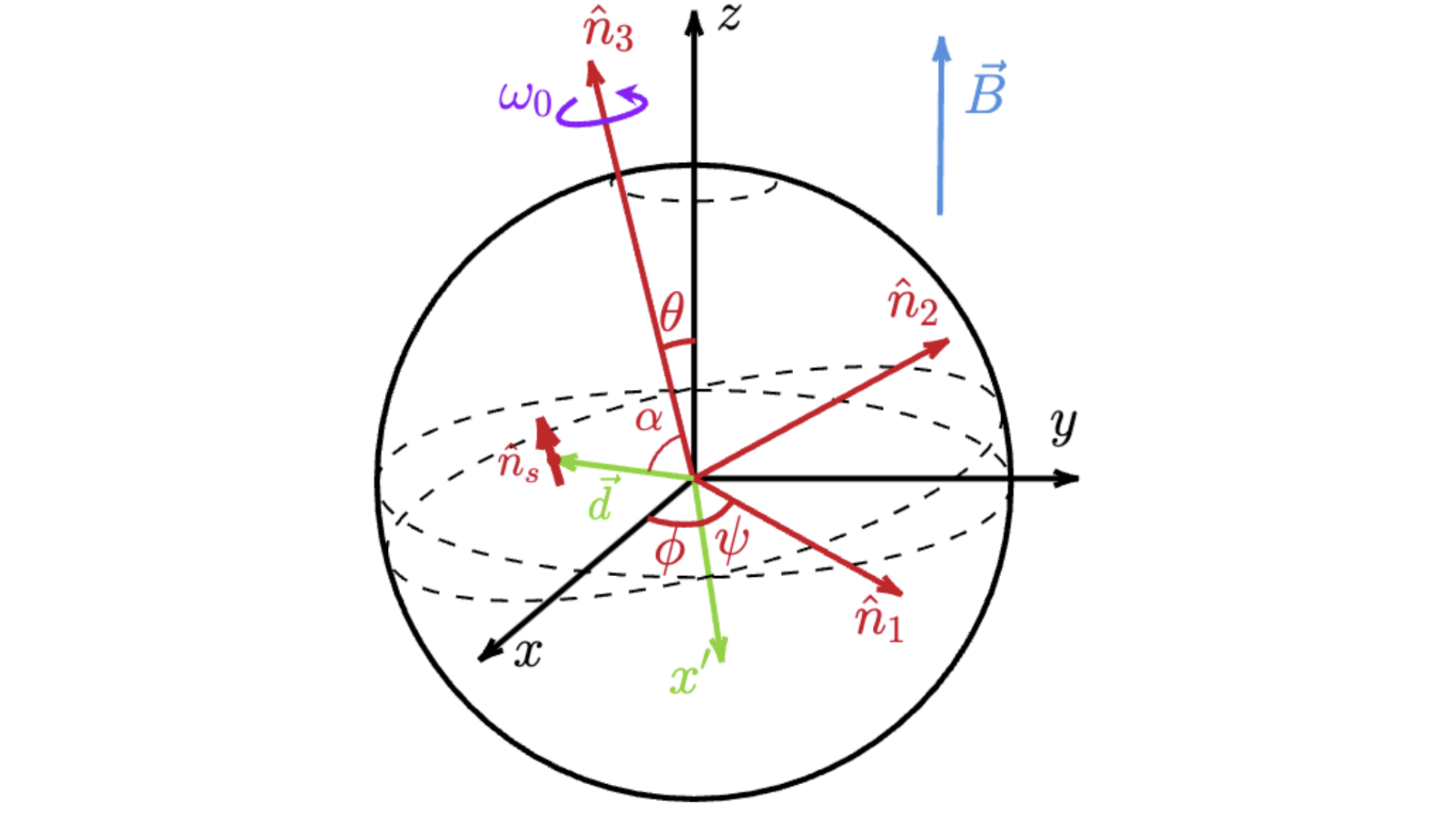}
\caption{The ND sphere model and the three Euler angles $\{\phi, \theta,\psi\}$ in the ZXY convention. We take the gauge that $\{\hat{n}_1,\hat{n}_2,\hat{n}_3\}$ are coincide with $\{x,y,z\}$ respectively when $\phi=\theta=\psi=0$. $\vec{d}$ represents the vector of the location of the NV spin, and the angle between $\hat{n}_3$ and $\vec{d}$ is labelled as $\alpha$. The external magnetic field is aligned along the $z$ and $x$-axis. However, we assume that the superposition is one dimensional and will be created {\it solely} along the $z$-axis; we set $x=0$ always. Our new proposition to mitigate the loss of spin contrast is that the ND is initially spinning around $\hat{n}_3$-axis with frequency $\omega_0$.}
\label{eulerangles}
\end{figure}


\section{Stability of the libration mode by imparting rotation along the NV axis}

We propose the following scheme to alleviate the fine-tuning of parameters for the spin contrast loss in the SGI setup. The key is to give the ND a sufficient initial angular momentum that points to the NV axis. The model we built is shown in Fig. \ref{eulerangles}, where the ND is considered a nanosphere, namely $I_1=I_2=I_3=I$, following~\cite{Japha:2022phw}.
The rotational motion is described by three Euler angles: 
\begin{itemize}
\item Precession angle is given by $\phi$. 

\item Libration/nutation angle is given by $\theta$. The dynamics of the libration mode are the most important for us. See Fig.~\ref{eulerangles}. The new protocol will be to impart external rotation at the beginning of the experiment along the NV axis, as shown in the figure.

\item Rotation angle is given by $\psi$. 
\end{itemize}

We study the general case where the NV center is not located at the center of mass (COM) of the ND. We take the following steps now.

\begin{itemize}

\item Initially, we levitate an ND sphere and use circularly polarized light to derive stable rotation of the ND with an angular velocity $\omega_0$ around the $\hat{n}_3$ axis, which is experimentally feasible\cite{friese1998optical, jin2024quantum,perdriat2023spin, kuhn2017optically, Afek:2021bua,reimann2018ghz}. According to the most recent experiment\cite{jin2024quantum}, optically induced rotation of levitated ND has been performed at high frequencies up to $20$~MHz. Then, the NV spin state is prepared as the superposition state $|s\rangle=(\lvert+1\rangle+\lvert-1\rangle)/\sqrt{2}$. 

\item The next step is to turn on the magnetic field $\vec{B}(z,x)$ at $t=0$, where the $\hat{z}$ direction should coincide as much with the initial ND rotation axis $\hat{n}_3$ as possible. A small initial angular mismatch $\theta_0$ between $\hat{n}_3$ and $\hat{z}$ can be tolerated. Thanks to the gyroscopic effect, the rotation axis of the fast-spinning ND is now stable, which means that the orientation of $\hat{n}_3$ and $\hat{n}_{s}$ can always be approximately aligned with the $\hat{z}$ axis. 

Therefore, in this case, when considering the spatial trajectory of the two SGI arms, one can safely assume that the NV spin orientation always keeps parallel to the magnetic direction $\hat{z}$. 
\end{itemize}

In this way, the spatial trajectories of the SGI model do not change significantly due to the gyroscopic stability of the ND.

Furthermore, we should ensure that the NV state $s=\pm1$ is stable during SGI. However, some special values of the rotation frequency $\omega_0$ possibly lead to the spin-resonant transition $s=\pm1\leftrightarrow 0$\cite{chen2019nonadiabatic,ma2021torque}, which should be avoided to maintain the spin state during the SGI process. First, we investigate the spin dynamics in a rotating ND to find the condition for the stability of spin. Then, we focus on the gyroscopic stability of the ND and the angular mismatch of the Euler angles.



\subsubsection{Evolution of Spin and Einstein-de Hass effect}\label{EdH}

In the Heisenberg picture, based on the Hamiltonian \eqref{hamiltonian}, the operator equations of motion for the NV spin and ND's rotation dynamics are given by\cite{Japha:2022phw}
\begin{equation}\label{dotS}
\dot{\textbf{s}}= \frac{\mu}{\hbar} (\textbf{B}\times \textbf{s}) + \frac{i}{\hbar}[D(\textbf{s}\cdot\hat{\textbf{n}}_s)^2,\textbf{s}]\, ,
\end{equation}
\begin{align}\label{dotL}
\dot{\textbf{L}}&= \frac{i}{\hbar} D[(\textbf{s}\cdot\hat{\textbf{n}}_s)^2,\textbf{L}]  \nonumber\\
&=D((\textbf{s}\cdot\hat{\textbf{n}}_s)(\textbf{s}\times\hat{\textbf{n}}_s)+(\textbf{s}\times\hat{\textbf{n}}_s)(\textbf{s}\cdot\hat{\textbf{n}}_s)) \nonumber \\
&=-i [D(\textbf{s}\cdot\hat{\textbf{n}}_s)^2,\textbf{s}]\, ,
\end{align}
\begin{align}\label{dotni}
    \dot{\hat{\textbf{n}}}_i=\frac{i}{\hbar}\sum_k\left[\frac{L_k^2}{I_k},\, \hat{\textbf{n}}_i\right] =\boldsymbol{\omega} \times \hat{\textbf{n}}_i + i\gamma \hat{\textbf{n}}_i \, ,
\end{align}
where $i=\{1,2,3\}$ labels the index in the rotating frame and $\gamma=\hbar\sum_k(1/2I_k)$. Then, we have
\begin{align}\label{sdot}
    \dot{s}_i &= \dot{\textbf{s}}\cdot \hat{\textbf{n}}_i + \textbf{s}\cdot \dot{\hat{\textbf{n}}}_i \\
    & = \frac{i}{\hbar}[D(\textbf{s}\cdot\hat{\textbf{n}}_s)^2,s_i] + \frac{\mu}{\hbar}\varepsilon_{ijk}B_js_k + \varepsilon_{ijk}s_j \omega_k + i\gamma s_i\, , \nonumber
\end{align}
where $\varepsilon_{ijk}$ is the Levi-Civita epsilon and the Einstein summation convention is used here. For massive objects with $m\gtrsim 10^{-17}$kg, the coefficient $\gamma\sim \hbar/I\lesssim 10^{-8}$Hz is negligible compared to $D\sim$GHz, $\mu B/\hbar\sim$MHz and $\omega\sim$ kHz-MHz. We can define an effective Hamiltonian to describe the spin evolution in the rotating frame
\begin{equation}
    H_s=D S_{\parallel}^2 + \mu \textbf{s} \cdot \textbf{B} - \frac{\hbar}{I}\textbf{s}\cdot \textbf{L}\,. 
\end{equation}
It can be verified that the Hamiltonian equation $\dot{s}_i=i[H_s,s_i]/\hbar$ is consistent with (\ref{sdot}). The effect of the spin-rotation coupling $\textbf{s}\cdot \textbf{L}$ is known as the Einstein-de Hass (EdH) effect~\cite{barnett1915magnetization,einstein1915experimental,izumida2022einstein}. For our SGI model, applying the basis $\{s_1,s_2,s_3\}$, the effective spin Hamiltonian is
\begin{equation}
    H_s= \begin{pmatrix}
\Delta_+ & We^{-i\omega_0t} & 0 \\
We^{i\omega_0t} & 0 & We^{-i\omega_0t} \\
0 & We^{i\omega_0t} & \Delta_- 
\end{pmatrix} \, ,
\end{equation}
where 
\begin{equation}
    \Delta_\pm \equiv D\pm\mu B{\rm cos}\theta_0\mp\hbar\omega_0 ,\ W \equiv \frac{\mu B {\rm sin}\theta_0}{\sqrt{2}} \, .
\end{equation}
In the resonant case $\hbar\omega_0\approx D\pm \mu B{\rm cos}\theta_0$, $s_3=+1$(or $s_3=-1$) state would degenerate with the $s_3=0$ state, which means that any small off-diagonal term $W$ would leads to Rabi oscillation between $s_3=\pm1$ and $s=0$ state. To mitigate the amplitude of the Rabi oscillation, we take the off-resonant condition $\hbar\omega_0\ll D\pm \mu B{\rm cos}\theta_0$. Then, we have $|\Delta_{\pm}|\gg W$, so the Rabi oscillation is negligible. 

Moreover, in the case $\hbar\omega_0\approx \mu B {\rm cos}\theta_0$, we have $\Delta_+\approx\Delta_-$, which degenerates the two spin states $s=\pm1$. It leads to the spin-flip transition between $s=+1$ and $s=-1$ with probability, which is known as Majorana flipping \cite{majorana1932atomi,inguscio2006pos,Marshman:2021wyk} and should also be avoided in SGI.  Therefore, we apply the following condition to the initial rotation frequency
\begin{equation}\label{constraint1}
    \hbar\omega_0\ll\mu B {\rm cos}\theta_0 \, ,\ {\rm and}\  \hbar\omega_0\ll D\pm \mu B{\rm cos}\theta_0 \, .
\end{equation}

Therefore, the spin components in the rotating frame are stable during the interferometry process, that is, $\dot{s}_i=0$.


\subsubsection{Equations of motion of the Euler angles}

To describe rotational dynamics, we define the rotational Lagrangian $\mathcal{L}=T-V$, where the rotational kinetic energy $T$ of the ND sphere in terms of Euler angles is given by~\cite{Landau}
\begin{equation}
T\equiv \frac{\textbf{L}^2}{2I} =\frac{I}{2}(\dot{\theta}^2+\dot{\phi}^2{\rm sin}^2\theta) +\frac{I}{2} (\dot{\phi}{\rm cos}\theta+\dot{\psi})^2 \, ,
\end{equation}
and the potential term, $V$, generates torque acting on ND's rotation. 

According to the equation (\ref{dotS},\ref{dotL},\ref{dotni}) and $\dot{s}_i=\dot{\textbf{s}}\cdot\hat{\textbf{n}}_i+\textbf{s}\cdot\dot{\hat{\textbf{n}}}_i=0$, we can get the evolution of ND's rotational angular momentum
\begin{align}\label{dotL2}
    \dot{L}_i&= \dot{\textbf{L}}\hat{\textbf{n}}_i + \textbf{L}\cdot \dot{\hat{\textbf{n}}}_i = \mu (\textbf{B}\times\textbf{s})\cdot\hat{\textbf{n}}_i -\hbar \dot{\textbf{s}}\cdot\hat{\textbf{n}}_i\nonumber + \textbf{L}\cdot \dot{\hat{\textbf{n}}}_i\\
    &= \varepsilon_{ijk} \left(\mu B_j s_k + \hbar s_j\frac{L_k}{I_k} + L_j\frac{L_k}{I_k}   \right)\, ,
\end{align}
where the last term vanishes because the shape of the ND we take is spherical, e.g. $\dot{\hat{\textbf{n}}}_i=0$. Therefore, we can see that there are two torques that affect the rotational dynamics of the ND. The physical origin is the stable condition $\dot{s}=0$, so that the NV spin vector is fixed in the direction of the NV axis, namely, $\textbf{s}=s\hat{n}_{s}$. Then, the mechanical rotation of the ND will rotate the magnetic moment of the NV spin, $\mu\textbf{s}$, and the angular momentum of the spin, $\textbf{S}$, implying that the NV spin causes the two torques in \eqref{dotL2}.

\begin{itemize}

\item The first torque arises from the Zeeman term in the Hamiltonian Eq.\eqref{hamiltonian} : $\mu \textbf{s}\cdot\textbf{B}=\mu s\, \hat{\textbf{n}}_{s}\cdot \textbf{B}$. 

\item The second torque comes from the EdH effect. Because of the rotation of the NV axis, the spin angular momentum rotates with the ND. So, the ND gives a torque to the NV spin angular momentum $\textbf{S}=\textbf{s}\hbar$, namely $\textbf{M}\equiv d\textbf{S}/dt=(\textbf{L}\times\textbf{S})/I$. Interactively, the ND is affected by a reaction torque $\textbf{S}\times \textbf{L}/I$ from the NV spin. 
This means that the NV spin can still affect the ND's rotation to some extent even when $\textbf{B}=0$ via the EdH effect~\cite{ma2021torque}.

\end{itemize}



In this paper, by taking the constraint Eq.\eqref{constraint1} on $\omega_0$, we work in a regime where the effect of EdH torque in \eqref{dotL2} is negligible , e.g. 
\begin{equation}
\frac{\hbar L}{I}=\hbar\omega_0\ll\mu B.
\end{equation}
Thus, in the $V$ term in the Lagrangian density ${\cal L}=T-V$, only the Zeeman term $\mu \textbf{s}\cdot\textbf{B}=\mu s B {\rm cos}\theta$ is important for our dynamics, independent of the angular momentum variables. Then, we take $V=\mu s B {\rm cos}\theta$ can define the canonical momentum of the three Euler angles
\begin{eqnarray}\label{momentum}
p_\theta &\equiv& \partial \mathcal{L}/\partial \dot{\theta}= I\dot{\theta} \, ,\nonumber\\
    p_{\phi} &\equiv& \partial\mathcal{L}/\partial \dot{\phi}=I (\dot{\phi}+\dot{\psi}\,{\rm cos}\theta)=L_z  \, ,\\ \nonumber
    p_\psi &\equiv& \partial\mathcal{L}/\partial \dot{\psi}= I (\dot{\psi}+\dot{\phi}\,{\rm cos}\theta)=L_3\, .
\end{eqnarray}
Then, the effective Hamiltonian $H_r$ reads
\begin{equation}\label{rotationalhamiltonian}
    H_r=\frac{p_\theta^2}{2I}+\frac{p_\psi^2}{2I} + \frac{(p_\phi-p_\psi{\rm cos}\theta)^2}{2I\, {\rm sin}^2\theta} 
     + \mu s B_{nv} {\rm cos}\theta\, ,
\end{equation}
where $B_{nv}=B_c+\tilde{\eta}(t)d\,({\rm cos}\alpha\,{\rm cos}\theta+{\rm sin}\theta\,{\rm sin}\alpha\,{\rm cos}\psi)$ denotes the magnetic field strength at the location of the NV center, and $B_c$ is the magnetic field strength at the COM.

Since $\eta d\ll B_c$, we assume that $B_{nv}\approx B_c$ for our analysis. Therefore, the canonical momentum $p_\psi$ and $p_\phi$ are conserved (e.g. no external magnetic force is acting along these directions):
\begin{equation}\label{conserved}
p_\phi=I\omega_0 {\rm cos}\theta_0 \, , \quad p_\psi=I\omega_0\, .
\end{equation}

\subsubsection{Stability of $s=\pm 1$ states in the SGI setup}

Substituting the conserved quantities Eq.~(\ref{conserved}) to Eq.~(\ref{rotationalhamiltonian}), we obtain the Hamiltonian for the evolution of $\theta$. To make the libration of the NV axis small, here we demand the small oscillation condition $\theta-\theta_0\ll\theta_0\ll1$. Therefore, we end up with the equation of motion (e.q.m) of $\theta$ (the libration mode), as follows:
\begin{equation}\label{eomtheta1}
\ddot{\theta} = - \left(\omega_0^2-\frac{\mu s B_c(t)}{I}\right)(\theta-\theta_0) +  \frac{\mu s}{I}B_c(t)\theta_0\, ,
\end{equation}
The above Eq.(\ref{eomtheta1}) implies that the orientation of the NV axis is stable for both $s=+1$ and $s=-1$ spin states as long as the requirement $\omega_0^2> \mu B_c/I$ is satisfied. 

\begin{figure}       
\centering
\includegraphics[width=\linewidth]{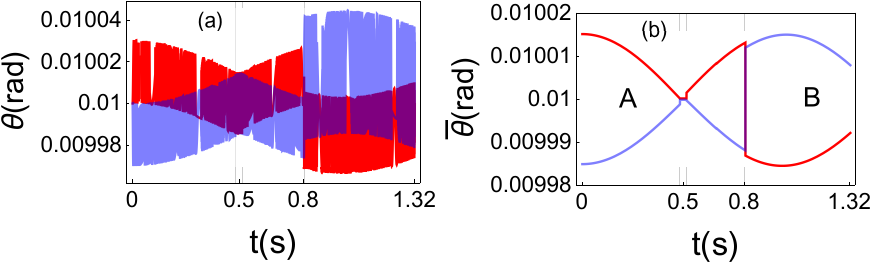}
\includegraphics[width=0.9\linewidth]{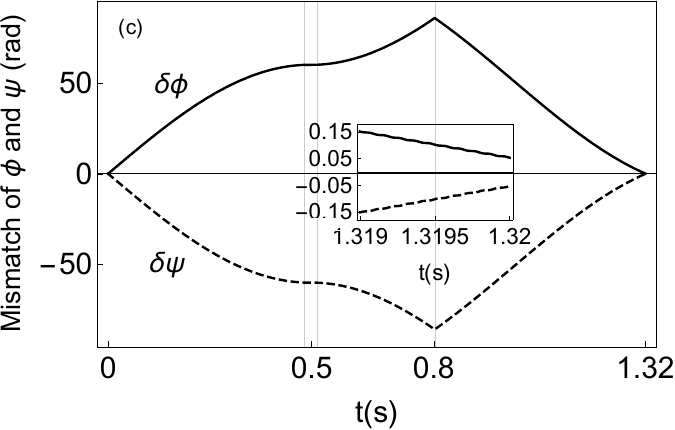}
\caption{The red and blue curve in Subfigure (a) represent the numerical time evolution of nutation mode $\theta(t)$ for the ND with initial spin state $s=+1$ and $s=-1$, respectively. Subfigure (b) shows the equilibrium position $\bar{\theta}(t)$ of the oscillation mode $\theta(t)$. Subfigure (c) shows the evolution of the procession angle difference $\delta\phi(t)$ and rotation angle difference $\delta\psi(t)$ between the two SGI arms. In this plot, we set $m=10^{-17}$kg, $B_0=100G$, $B_1=1.0G$, $\eta=0.45G/\mu m$. Note that following \cite{Marshman:2021wyk}, there are four time stages and we set $\tau_1=0.482s$, $\tau_2=0.514s$, $\theta_0=0.01$, $\omega_0=2\pi\times 10$kHz, $d=10$nm, $\alpha=\pi/6$. We reverse the NV spin state between $\lvert+1\rangle$ and $\lvert-1\rangle$ at $\tau_3=0.8022s$. The spatial trajectory of the two SGI arms closed at $\tau_4=1.320s$. Vertical lines show these time stages here, except $\tau_4=1.32s$.}
    \label{fig2}
\end{figure}

\begin{figure}   
\centering

\includegraphics[width=0.9\linewidth]{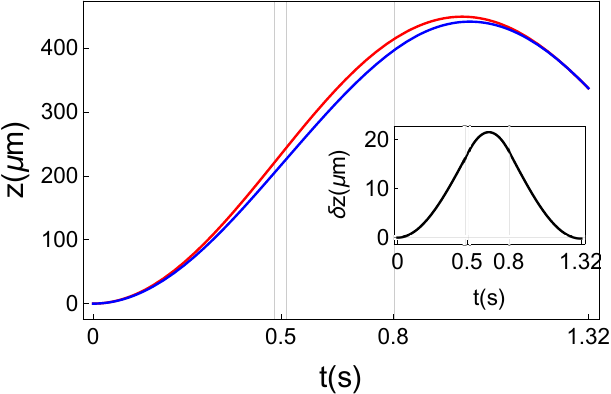}

    \caption{The red curve and blue curve represent the numerical simulation for spatial trajectories $z(t)$ of the ND with initial NV spin state $s=+1$ and $s=-1$, respectively. In this plot, we set the same parameters in Fig. \ref{fig2}. The maximum spatial superposition size $\delta z$ is about $20\mu$m. The initial rotation with $\omega_0=2\pi\times10$kHz ensures the stability of the orientation of the NV axis, namely the angle $\theta$ is always small, see Fig. \ref{fig2}(a). Therefore, the spatial trajectories of the SGI can be maintained as the original design in Ref. \cite{Marshman:2021wyk}. Following \cite{Marshman:2021wyk}, there are four time stages and we set $\tau_1=0.482s$, $\tau_2=0.514s$, $\theta_0=0.01$, $\omega_0=2\pi\times 10$kHz, $d=10$nm, $\alpha=\pi/6$. We reverse the NV spin state between $\lvert+1\rangle$ and $\lvert-1\rangle$ at $\tau_3=0.8022s$. The spatial trajectory of the two SGI arms closed at $\tau_4=1.320s$. Vertical lines show these time stages here, except $\tau_4=1.32s$}
    \label{ztpm}
\end{figure}

Further, to ensure the small oscillation condition, we need a sufficiently large initial angular velocity that satisfies 
\begin{equation}
    \omega_0^2\gg \frac{\mu B_c}{I} \, .
\end{equation}
Therefore, recall the constraint \eqref{constraint1}, the order of magnitude for parameters in the SGI experimental setup can be concluded as follows:
\begin{equation}
\sqrt{\frac{\mu B_c}{I}} \ll \omega_0\ll \frac{\mu B_c}{\hbar}\, ,\  {\rm and}\ \omega_0\ll \frac{D\pm\mu B_c}{\hbar}\, .
\end{equation}
For the experimental parameters set in Fig. \ref{fig2}, the order of magnitude of $\sqrt{\mu B_c/I}$ is approximately $2\pi\times10^2{\rm Hz}$, $\mu B_c/\hbar\sim 5-300{\rm MHz}$, and $D/\hbar=2\pi\times2.87$GHz. So, the value of $\omega_0$ can be set in the order of $\omega_0\sim{\rm kHz}-{\rm MHz}$. Fig. \ref{ztpm} shows the spatial trajectories of the SGI model for a suitable value of $\omega_0$.


\subsubsection{Evolution of the libration mode and the final mismatch}

We need to compute the evolution of the libration mode $\theta(t)$, and study the overlap of the libration mode (for the two arms of the SGI) upon the completion of the interferometer.

For long-duration SGI models, the variation of $B_c$ in Eq.~(\ref{eomtheta1}) is slow compared to the fast oscillation of $\theta$. So, the libration mode $\theta(t)$ shows a fast harmonic oscillation with frequency $$\omega=\sqrt{\omega_0^2-\mu sB_c/I}\approx\omega_0$$ and an adiabatically evolving equilibrium position $\bar{\theta}(t)$, illustrated in Fig. \ref{fig2}(a,b), where $\bar{\theta}$ is obtained from Eq.~(\ref{eomtheta1}) as 
\begin{equation}\label{thetabar}
\bar{\theta}(t) \approx \theta_0 + \frac{\mu s }{I\omega_0^2} B_c(t) \theta_0 \, .
\end{equation}
Here, we take the ground state of the harmonic trap as the initial state of the libration mode, i.e., $\theta(0)=\theta_0$ and $\dot{\theta}=0$. 

The amplitude of the libration mode $\theta(t)$ can be estimated as $A_\theta(0)=\mu B_0\theta_0/I\omega_0^2$ during $0\le t<\tau_3$. At $t=\tau_3$, the NV spin is flipped, then the equilibrium position $\bar{\theta}$ changes rapidly, disturbs the oscillation, and the libration amplitude is somewhat enhanced. Nevertheless, the final libration amplitudes of the two SGI arms are still on the same order of magnitude of $\mu B_0\theta_0/I\omega_0^2$. 

Therefore, the final mismatch of $\theta$, denoted as $\delta\theta$ (between the two arms of the SGI at the time of closing the one-loop interferometer), should also be in the same order. More precisely, the maximum of $\delta\theta$ can be estimated by (see Appendix \ref{appB} for the derivation)
\begin{equation}\label{mismatchtheta}
\delta\theta(\tau_4) \lesssim \frac{8\mu B_0\theta_0}{I\omega_0^2}  \, .
\end{equation}
The mismatch of libration angle $\theta$ is inevitable, but the large $\omega_0$ can highly reduce it to a small value of $\theta_0$. For the case of $\omega_0=10$~kHz, $\theta_0=0.01$ and $B_0=100$~G, the mismatch of $\theta$ on the order of $10^{-5}$, incredibly small.


\subsubsection{Evolutions of Precession and rotation angles}

For the precession $\phi(t)$ and the rotation mode $\psi(t)$, according to Eq.~(\ref{momentum}) and (\ref{conserved}), the e.q.m are
\begin{align}
\dot{\phi}&=\frac{\omega_0}{{\rm sin}^2\theta_0}({\rm cos}\theta_0-{\rm cos}\theta)\approx \frac{\omega_0}{\theta_0}(\theta-\theta_0) \,, \\
\dot{\psi}&=\frac{\omega_0}{{\rm sin}^2\theta_0}(1-{\rm cos}\theta_0{\rm cos}\theta)\approx -\frac{\omega_0}{\theta_0}(\theta-\theta_0) \,,
\end{align}
where we use the condition $\theta-\theta_0\ll\theta_0\ll1$ Rad for our setup (a small angle approximation for $\theta$). 

We can compute the mismatch in the evolution of $\phi$ and $\psi$ during the creation and annihilation of the spatial superposition after the completion of the one-loop interferometry. The mismatch of classical trajectory of $\phi(t)$ is given by
\begin{equation}\label{mismatchphi}
\delta\phi (\tau_4)\approx\frac{\omega_0}{\theta_0} \int_0^{\tau_4} dt\, (\bar{\theta}_L-\bar{\theta}_R) = \frac{\omega_0}{\theta_0} (\Sigma_A-\Sigma_B)\, ,
\end{equation}
where $\bar{\theta}_L$ and $\bar{\theta}_R$ represent the red and blue curves shown in Fig. \ref{fig2}(b), and $\Sigma_A$ and $\Sigma_B$ labels the area of the region $A$ and $B$ respectively. The mismatch of $\psi$ is given by $\delta\psi=-\delta\phi$. 

Generally, the area $\Sigma_A$ and $\Sigma_B$ are unequal. However, one can regulate the trajectory of $\bar{\theta}(t)$ by optimizing the experimental parameters, such as $B_1$ and $B_0$, to obtain $\Sigma_A\approx\Sigma_B$. For example, as illustrated in Fig. \ref{fig2}(c), the parameters used in Fig. \ref{fig2} result in a nearly vanishing mismatch of $\phi$ and $\psi$, in which $|\delta\phi|=|\delta\psi|\approx0.05$. Remarkably, from Eqs.~(\ref{thetabar},\ref{mismatchphi}), we have $\delta\phi \propto \omega_0^{-1}$. So, the mismatch $\delta\phi$ and $\delta\psi$ can be further reduced by applying a larger $\omega_0$.


\section{spin contrast}

The spin contrast of the interferometer, denoted by $C$, is defined by the recombination between the quantum wave function of the two interferometer arms,
$C=|\langle\Psi_{L}(\tau)|\Psi_{R}(\tau)|$, denoted here by left (L) and right (R). Here,
$\tau$ is the time scale of completing the interferometer, in our SGI model, $\tau =\tau_4$, see the discussion in Sec.\ref{ModelofSGI}.

Let us figure out the spin contrast loss due to the mismatch of rotational d.o.f.. We consider the initial wave packets with the Gaussian profile of the three rotational d.o.f. The initial quantum uncertainty of momentum is denoted as $\Delta p_\psi$ and $\Delta p_\phi=\Delta p_\psi{\rm cos}\theta_0$. The initial quantum state of the libration mode $\theta$ is considered a coherent state of the harmonic trap with frequency $\omega_0$. 

The wave packets of $\theta$ in case $s=\pm 1$ keep coherent states during evolution, since they are both confined in traps with frequency $\omega_0$. The overlap between two symmetric Gaussian wave packets can be estimated by applying the semiclassical approach\cite{SGI_experiment,japha2022role}, namely
\begin{equation}\label{semiclassical}
    C \propto {\rm exp}\left( -\frac{1}{2}\frac{\delta\theta^2}{l_\theta^2}- \frac{1}{2}\frac{\delta p_\theta^2}{l_p^2} \right) \, ,
\end{equation}
where $\delta\theta$ and $\delta p_\theta$ are the classical mismatch of $\theta$ and $p_\theta$, respectively. $l_\theta\equiv \hbar/ \Delta p_\theta$ and $l_p\equiv \hbar/ \Delta \theta$ are the spatial coherence length and momentum coherence width for the $\theta$ mode, respectively, in which $\Delta\theta$ and $\Delta p_\theta$ denote the quantum uncertainty of $\theta$ and the momentum $p_\theta$. We can see that the spin contrast is determined by the mismatch of the two classical trajectories. Moreover, there is no spin coherence loss caused by the shape mismatch of the left and right wave packets, while the scheme using the superposition of $s=0,-1$ states suffers from this problem\cite{Japha:2022phw}. According to the analysis of the classical dynamics of $\theta$, the order of magnitude of $\delta\theta$ and $\delta p_\theta$ can be estimated by $\delta\theta \sim \mu B_0\theta_0/(I\omega_0^2)$ and $\delta p_\theta \sim I\omega_0\delta\theta\sim \mu B_0\theta_0/\omega_0$. For a coherent state, the quantum uncertainty reads $\Delta\theta=\sqrt{\hbar/(2I\omega_0)}$ and $\Delta p_\theta \sim \sqrt{\hbar I\omega_0/2}$. Therefore, By the semiclassical method \eqref{semiclassical}, the spin contrast loss due to the mismatch of the libration mode is roughly
\begin{equation}\label{Ctheta}
    C \propto {\rm exp}\left( -\frac{N\mu^2B_0^2\theta_0^2}{\hbar I \omega_0^3}  \right) \, ,
\end{equation}
where $N$ is a coefficient. 

Similarly, the spin contrast decay due to the mismatch $\delta \phi$ and $\delta\psi$ can also be estimated by using the semi-classical approach \eqref{semiclassical}, i.e.
\begin{equation}\label{Cphipsi}
    C\propto {\rm exp} \left(-\frac{1}{2}\delta\phi^2 \frac{\Delta p_\phi^2}{\hbar^2} -\frac{1}{2}\delta\psi^2 \frac{\Delta p_\psi^2}{\hbar^2} \right) \, ,
\end{equation}
in which we have used the coherent length $l_\phi=\hbar/\Delta p_\phi$, $l_\psi=\hbar/\Delta p_\psi$ and the classical mismatch of momentum $\delta p_\phi=\delta p_\psi=0$ because of the momentum conservation of the $\phi$ and $\psi$ mode.  

\begin{figure}[ht]       
\centering
\includegraphics[width=0.8\linewidth]{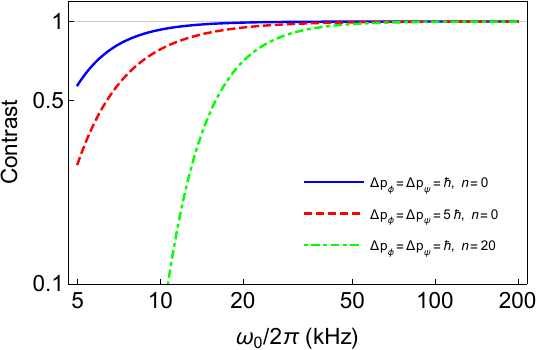}
\caption{The bound of the spin contrast as a function of the initial angular velocity $\omega_0$ for different quantum uncertainties of the angular momentum $I\omega_0$. In this plot, we take the same parameters as in Fig. \ref{fig2}. The dotted-dished curve shows the contrast in the finite temperature case, where $n=k_BT/(\hbar\omega_0)$. We can see that, in the case of $\omega_0=2\pi\times50$kHz, the contrast is still close to 1 when $n=20$ ($T\sim0.05$mK, which is achievable based on recent rotational cooling technologies~\cite{schafer2021cooling}).}
\label{contrastfig}
\end{figure}

Beyond the semi-classical approach described above, we derive the spin contrast by the quantum approach(see Appendix. \ref{appC} and \ref{appD}). The lower bound of the final spin contrast can be evaluated by
\begin{equation}\label{contrast}
C > exp\left(-\frac{1}{2}\delta\phi^2 \frac{\Delta p_\phi^2}{\hbar^2} -\frac{1}{2}\delta\psi^2 \frac{\Delta p_\psi^2}{\hbar^2} -\frac{16\mu^2B_0^2\theta_0^2}{\hbar I \omega_0^3} \right) \, ,
\end{equation}
which is illustrated in Fig. \ref{contrastfig}.

The result Eq.\eqref{contrast} is consistent with the semi-classical estimation Eq.\eqref{Ctheta} and \eqref{Cphipsi}. As we can see, the final SGI contrast is protected by large values of $\omega_0$, which reduces all mismatches $\delta\phi=-\delta\psi\propto\omega_0^{-1}$, and $\delta\theta\propto \omega_0^{-2}$. 
According to the recently reported experiment about rotating levitated ND \cite{jin2024quantum}, the line width of the rotational frequency $\omega_0$ is measured on the order of $10^{-4}$Hz for ND with raduis $264$nm ($m\approx3\times 10^{-16}$kg) when $\omega_0=100$kHz, which corresponds to uncertainty $\sigma= I\Delta\omega\approx7\hbar$. Hence, current ND rotation control techniques can be applied to protect the spin contrast in future SGI experiments. 

We can extend our estimation to finite temperature $T$, the initial state of the nutation mode can be considered as $\rho_{th}=\int d^2\alpha P(\alpha)|\alpha\rangle\langle\alpha|$, where $|\alpha\rangle$ represents coherent states, and $P(\alpha)$ is the distribution function of the thermal state. The lower bound of the spin contrast in finite temperature case is given by (see the derivation in Appendix \ref{appE}):
\begin{equation}\label{contrastthermalbound0}
C_{th} >  exp \left[ -\frac{\delta\phi^2\Delta p_\phi^2}{2\hbar^2} - \frac{\delta\psi^2\Delta p_\psi^2}{2\hbar^2} - \frac{16(1+2n)\mu^2B_0^2\theta_0^2}{\hbar I \omega_0^3} \right] \
\end{equation}
Thermal noise causes a decrease in the overlap between the two nutation states of the two SGI arms, which can be qualified as the $(1+2n)$ factor that multiplies the last term in the exponential of (\ref{contrast})(see Appendix \ref{appE}), where $n=k_BT/(\hbar\omega_0)$. An example with $n=20$ is shown in Fig. \ref{contrastfig}.


\section{Conclusion}

We proposed that, without changing the SGI scheme, giving an appropriate initial angular momentum along the superposition direction to the massive object can cure the interferometry contrast loss induced by the rotational dynamics of the object. Because gyroscope stability is a universal property of fast-spinning objects, the method is scheme-independent. It can be applied to other SGI models and other materials of nanoparticles with an embedded spin defect. Remarkably, the quantum wave packets of the two SGI arms are symmetric in our model, which avoids contrast loss induced by the asymmetric evolutions of quantum uncertainty (see \cite{Japha:2022phw} and Appendix \ref{appA}). In other words, only contrast loss is caused by the classical mismatch of the two angular trajectories. Moreover, the process does not require fine-tuning of the experimental parameters. 
Furthermore, there is no constraint on the location of the NV center in ND. The only assumption is that the shape of ND is spherically symmetric. If the ND is not a sphere, we require that the NV axis is oriented to the principal axis with the maximum or minimum inertia moment because objects are rotationally unstable around the axis of intermediate inertia moment\cite{ma2020quantum, stickler2021quantum}. A detailed analysis is pending for a future paper. Last but not least, the rotational effects of a nanoparticle are also known to ameliorate the intrinsic dipole moment\cite{Afek:2021bua}.


\section*{Acknowledgments}

T.Z is supported by the China Scholarship Council (CSC). AM and SB's research is funded in part by the Gordon and Betty Moore Foundation through Grant GBMF12328, DOI 10.37807/GBMF12328. This material is based upon work supported by Alfred P. Sloan Foundation under Grant No. G-2023-21130.
The authors thank T.H. Taminiau for helpful discussions on the coherence time of NV spins and nuclear spins in diamond, and also thank Yonathan Japha and Ron Folman for the insightful discussions regarding their findings.

\appendix

\section{Motion of a nanodiamond without  rotation}\label{appA}
In this part, we show that, in the case when ND is initially static, namely $\omega_0=0$, it is hard to obtain a large spin contrast of SGI with massive objects and long duration, and the spatial superposition size would be half of the original SGI designs.

\begin{figure}[ht]       
\centering
\includegraphics[width=0.9\linewidth]{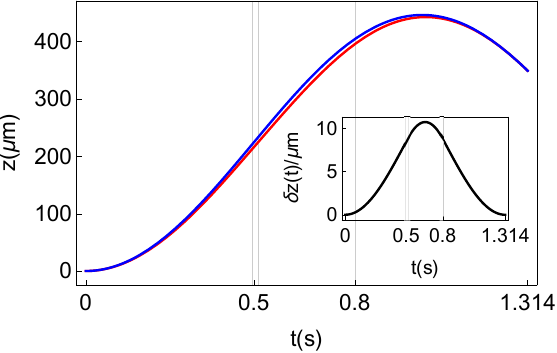}
\caption{The red curve and blue curve represent the spatial trajectories $z(t)$ of the ND with initial NV spin state $s=-1$ and $s=0$, respectively. We reverse the NV spin state between $\lvert0\rangle$ and $\lvert-1\rangle$ at $\tau_3=0.800s$. The position and translational momentum of the two arms can match at $\tau_4=1.314s$. In this plot, we set $m=10^{-17}kg$, $B_0=100G$, $B_1=1.0G$, $\eta=0.45G/\mu m$, $\tau_1=0.494s$, $\tau_2=0.513s$, $d=10 nm$, $\alpha=\pi/6$. The maximum of the superposition size $\delta z(t)$ is about 10$\mu m$. The scheme has a half superposition size $\delta z$ compared with the scheme using the spin superposition state of $s=+1$ and $s=-1$, see Fig. \ref{ztpm}. }
    \label{ztnz}
\end{figure}

\begin{figure}[ht]       
\centering
\includegraphics[width=\linewidth]{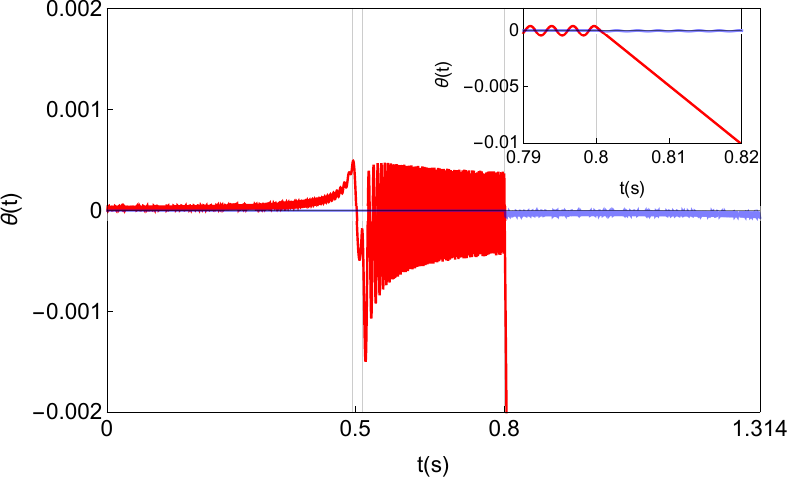}
\caption{The red curve and blue curve represent the trajectories $\theta(t)$ for the ND with initial NV spin state $s=-1$ and $s=0$, respectively. The parameters we set are the same as in Fig. \ref{ztnz}. Before $t=\tau_3=0.800$s, the ND with $s=-1$ state shows libration with frequency $\omega=\sqrt{\mu B_c/I}$. After $t=\tau_3$, the spin turns to $s=0$ state so that the ND will freely rotate, which leads to a large final mismatch $\delta\theta(\tau_4)\sim 0.3$.}
    \label{figa2}
\end{figure}

The Zeeman term $\mu sB{\rm cos}\theta$ in the ND's Hamiltonian implies that the ND with $s=-1$ state is rotationally stable when $\theta=0$, conversely, ND with $s=+1$ is unstable \cite{Japha:2022phw, japha2022role}.
To avoid a large rotation of angle $\theta$, we use the NV spin state superposition of $s=0$ and $s=-1$ state to create a spatial superposition of ND. From the Hamiltonian [Eq.(1) of the main text] of the ND, the equations of spatial and rotational motion are
\begin{equation}\label{ryanxequ}
    m\ddot{z}=-\frac{-\chi_{\rho}m}{\mu_0}B_c\tilde{\eta} - \mu s \tilde{\eta} {\rm cos}\theta \, ,
\end{equation}
\begin{align}\label{ryanthetaequ}
    \ddot{\theta}&=\mu s[ B_{nv}\,{\rm sin}\theta + \, \tilde{\eta}d\, {\rm sin}(\theta+\alpha){\rm cos}\theta ]\nonumber\\
    &\approx s\, \omega^2(t) \left( \theta - \frac{\tilde{\eta}d\,{\rm sin}\alpha}{B_c} \right) \, ,
\end{align}
where $B_{nv}=B_c+\tilde{\eta}d {\rm cos}(\theta+\alpha)$ and the libration frequency $\omega(t)=\sqrt{\mu B_c(t)/I}$. Assuming that the spatial and angular initial state is prepared as ground states in the traps of $z$ and $\theta$ d.o.f., namely $z(0)=\dot{z}(0)=\theta(0)=\dot{\theta}(0)=0$, the numerical solutions of the equation of motions for the two SGI paths are illustrated in Fig. \ref{ztnz},~\ref{figa2}.


The two spatial trajectories can overlap when we choose appropriate values of $\tau_1$, $\tau_2$, $\tau_3$ and $\tau_4$. However, the superposition size will be halved compared to the same SGI scheme with $s=\pm 1$. Even worse, the angular trajectories of the two paths can not finally recombine well, because the trajectory with initial NV state $s=-1$(see the red curve in Fig. \ref{figa2} deviates a lot from $\theta=0$ at $t=\tau_4$ in the context of the magnetic field profile chosen in the main text. To estimate the final mismatch of the trajectories of $\theta$, we describe the oscillation of $\theta$ as $A{\rm cos}( \omega (t-\tau_3)+\gamma_\theta)$ when $t\rightarrow \tau_3$, where the amplitude can be estimated by
$A\approx \eta d {\rm sin}\alpha/B_1$, $\omega=\sqrt{\mu B_c(\tau_3)/I}$, and $\gamma_\theta$ represents the oscillation phase at $t=\tau_3$. Therefore, at $t=\tau_3$, we have $\dot{\theta}(\tau_3)=A\omega {\rm sin}\gamma_\theta$. Then, because the NV spin state is flipped to the $s=0$ state, there is no torque acting on the ND, so the ND rotates with constant angular velocity $\dot{\theta}(\tau_3)$ until $t=\tau_4$. The final mismatch of $\theta$ reads:
\begin{align}\label{adeltatheta}
    \delta\theta (\tau_4) &\approx \dot{\theta}(\tau_3) \times(\tau_4-\tau_3) \nonumber\\
    &\approx \frac{\eta d {\rm sin}\alpha}{B_1} \sqrt{\frac{\mu B_0}{I}}(\tau_4-\tau_3) {\rm sin}\gamma_\theta\, ,
\end{align}
where we have used $B_c(\tau_3)\approx B_0$. And the mismatch of angular momentum is estimated by
\begin{equation}\label{adeltap}
    \delta p_\theta (\tau_4) \approx I\frac{\eta d {\rm sin}\alpha}{B_1} \sqrt{\frac{\mu B_0}{I}} {\rm sin}\gamma_\theta\, .
\end{equation}

We assume that the wave packet of $\theta$ is initially a coherent state of the harmonic oscillation mode. At $t=\tau_3$, the quantum uncertainties are $\Delta\theta=\sqrt{\hbar/2I\omega}$ and $\Delta p_\theta=\sqrt{\hbar I\omega/2}$. At $t=\tau_4$, the angular momentum uncertainty remains $\Delta p_\theta=\sqrt{\hbar I\omega/2}$ but the angular uncertainty expands to $\Delta\theta(\tau_4)\approx \hbar (\tau_4-\tau_3)/(2m \Delta\theta(\tau_3))$. Therefore, the coherent length of the Gaussian wave packet in angular and angular momentum space can be estimated by\cite{SGI_experiment, schaff2014interferometry, japha2022role}
\begin{equation}
    \lambda_\theta\approx \frac{\hbar}{\Delta p_\theta}\, ,\ 
    \lambda_p\approx \frac{\hbar}{\Delta \theta} \, ,
\end{equation}
and the spin contrast for two Gaussian wave packets with the same coherent length can be estimated by
\begin{equation}\label{semiclassicalcontrast}
    C\approx exp\left[-\frac{1}{2}\left( \frac{\delta\theta^2}{\lambda_\theta^2} + \frac{\delta p_\theta^2}{\lambda_p^2} \right)\right]\, .
\end{equation}
For the SGI example shown in Fig. \ref{ztnz},\ref{figa2}, the coherent length is about $\lambda_\theta\sim 0.002$ and $\lambda_p\sim I\times 0.003/s$. It means that the classical mismatch $\delta\theta$ and $\delta\dot{\theta}$ should be on or smaller than the order of $10^{-3}$Rad and $10^{-3} {\rm Rad/s}$, respectively, to obtain enough contrast. However, as shown in Fig. \ref{figa2}, when $d=10$nm and $\alpha=\pi/6$, the final mismatch is about $\delta\theta\approx 0.3$ and $\delta\dot{\theta}\sim 0.6/s$, which are much greater than the coherent length of the wave packet. So, the final contrast is almost zero. To obtain enough contrast, according to equation (\ref{adeltatheta},\ref{adeltap}), one can take the parameters $d$ and $\alpha$ satisfying $|d\,{\rm sin}\alpha|\lesssim 0.05$nm, for the parameters set in Fig. \ref{ztnz},\ref{figa2}, which is a strong constraint for the crystal structure. 

Remarkably, even if $d\, {\rm sin}\alpha =0$, i.e. the mismatch $\delta\theta$ and $\delta p_\theta$ of classical trajectories vanish, the contrast loss is still unavoidable because of the asymmetric quantum evolution of the wave packet of the two SGI arms, which is named the "quantum uncertainty limit" of spin contrast in \cite{Japha:2022phw}. The following form can describe the Gaussian wave function for the two SGI arms
\begin{equation}
    \Psi_a = \left(\frac{1}{2\pi \sigma_a^2}\right)^{\frac{1}{4}} exp\left( -\frac{\theta^2}{4\sigma_a^2} + \frac{i \beta_a \theta^2}{2} \right)\, ,
\end{equation}
where $\sigma_a(t)$ is the quantum uncertainty of the Gaussian wave packet and $a\rightarrow \{L,R\}$ labels the left and right arm of the SGI. According to the Schrodinger equation, the variable $\beta(t)=I\dot{\sigma}_a/(\hbar\sigma_a)$ is related to the expansion rate of the Gaussian wave packet\cite{japha2021unified}. Therefore, the contrast is
\begin{align}\label{contrastquantummismatch}
    &C\equiv \int_{-\infty}^{\infty} d\theta\, \Psi_L \Psi_R^* \nonumber\\
    &= \left[ \left(1+ \frac{(\sigma_L-\sigma_R)^2}{2\sigma_L\sigma_R}\right)^2 + (\beta_L-\beta_R)^2\sigma_L^2\sigma_R^2 \right]^{-\frac{1}{4}} \nonumber\\
    &= \left[\left( 1+ \frac{(\sigma_L-\sigma_R)^2}{2\sigma_L\sigma_R} \right)^2 + \frac{I^2}{\hbar^2}(\dot{\sigma}_L\sigma_R-\dot{\sigma}_R\sigma_L)^2 \right]^{-\frac{1}{4}} \, .
\end{align}
Let us label the SGI path with $s=-1$ initial state as the left arm. During $t<\tau_3$, the Gaussian wave packet is trapped in harmonic potential so that the uncertainty remains $\sigma_L(t)\approx \sigma_L(0)=\sqrt{\hbar/(2I\omega(0))}\equiv\sigma_0$. When $t>\tau_3$, due to $s=0$, the Gaussian wave packet is not affected by the harmonic trap so that the uncertainty of $\theta$ will expand, satisfying
\begin{equation}\label{sigmaL}
    \sigma_L(t)\approx\frac{\hbar}{2 I \sigma_0}(t-\tau_3)\, ,\quad t>\tau_3\, .
\end{equation}
For the right arm, during $t<\tau_3$, the spin state is $s=0$, so the wave packet is free to expend. When $t=\tau_3$, the uncertainty increases to $\sigma_R(\tau_3)\approx \hbar\tau_3/(2I\sigma_0)$. After $t=\tau_3$, the spin state turns to $s=-1$, then the evolution of the wave function will change from expansion to oscillation. Under the adiabatic approximation $\dot{\omega}\ll \omega^2$, the solution of uncertainty oscillation can take the form as(see more calculations in \cite{japha2021unified})
\begin{align}\label{sigmaR}
    \sigma_R(t)&\approx \sigma_R(\tau_3)\sqrt{{\rm cos}^2\gamma_\sigma+\frac{\hbar^2}{4I^2\omega^2\sigma_R^4(\tau_3)}{\rm sin}^2\gamma_\sigma} \nonumber\\
    &\approx \frac{\hbar\tau_3}{2I\sigma_0} |{\rm cos}\gamma_\sigma| \, ,\quad t>\tau_3 \, ,
\end{align}
where $\gamma_\sigma\equiv \int_{\tau_3}^t\omega(t')dt'$ represents the phase of the uncertainty oscillation and we have used $\sigma_R(\tau_3)\gg \sqrt{\hbar/(2I\omega(t))}\approx \sigma_0$ at the second step because $\omega \tau_3\gg 1$.
\begin{figure}[ht]       
\centering
\includegraphics[width=0.99\linewidth]{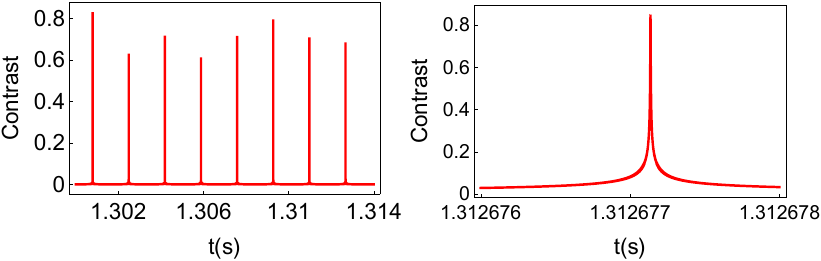}
\caption{Left panel: The evolution nearing $t=\tau_4$ of the contrast induced by the two asymmetric Gaussian wave packets of the SGI model. In this plot, we set $d=0$ so there is no classical mismatch between the two Gaussian wave packets. Other parameters are set the same as in Fig. \ref{ztnz}. We can see that there are periodically occurring peaks of contrast, and the frequency is about $2\omega(\tau_4)\sim 2\pi\times 300$Hz. Right panel: Enlarged plot for a contrast peak. The contrast peaks are very sharp, so the time scale of fine-tuning to obtain enough contrast should be much smaller than the order of microseconds.} 
    \label{contrastpeaks}
\end{figure}

Therefore, applying (\ref{sigmaL},\ref{sigmaR}) to (\ref{contrastquantummismatch}), we
get the evolution of the overlap of the two asymmetric wave packets, as illustrated in Fig. \ref{contrastpeaks}. The contrast oscillates with frequency $2\omega$(also shown in other models, e.g. \cite{Japha:2022phw, japha2022role}). By fine-tuning the final spin measurement time $\tau_4$, it is possible to reach a contrast peak when $t=\tau_4$. However, as shown in the right panel of Fig. \ref{contrastpeaks}, the precision of fine tuning should even be at the order of nanoseconds, which is much smaller than the SGI time scale. Moreover, some noise, such as the fluctuation of the magnetic field, can affect the oscillation of the wave packet's uncertainty, which causes unpredictable shifts of the location of contrast peaks, making fine-tuning more difficult. By the way, considering the thermal noise of the initial state, the contrast will be further reduced by one or more orders of magnitude~\cite{Japha:2022phw, japha2022role}.


\section{Classical evolution of the nutation mode}\label{appB}

The evolution of $\theta(t)$ is governed by the equation
\begin{equation}
    \ddot{\theta} = - \left(\omega_0^2-\frac{\mu s B_c}{I}\right)(\theta-\theta_0) +\frac{\mu s B_c\theta_0}{I}\, .
\end{equation}
The equation can be rewritten as
\begin{equation}\label{eomtheta2}
    \ddot{\theta} = - \omega^2(t)(\theta-\bar{\theta}(t))\, ,
\end{equation}
where
\begin{equation}
    \omega^2(t)\equiv \omega_0^2 - \frac{\mu s B_c(t)}{I}\approx\omega_0^2\, ,
\end{equation}
\begin{equation}\label{appendixbartheta}
    \bar{\theta}(t)\equiv \theta_0 + \frac{\mu sB_c(t)\theta_0}{I\omega_0^2-\mu s B_c(t)}\approx \theta_0 + \frac{\mu sB_c(t)\theta_0}{I\omega_0^2} \, .
\end{equation}
For SGI models with long timescale $\sim 1s$, the variation of $B_c(t)$ is much slower than the oscillation of $\theta(t)$ with frequency $\omega_0\sim$kHz-MHz. So, equation (\ref{eomtheta2}) can be regarded as an equation of a harmonic oscillator with adiabatically changing parameters. Therefore, we can take the ansatz for the solution the oscillation
\begin{equation}
    \theta(t)=A_\theta {\rm cos}(\omega t) +\bar{\theta} \, .
\end{equation}
Assuming the initial condition $\theta(0)=\theta_0$ and $\dot{\theta}(0)=0$, we get the initial amplitude of the nutation mode
\begin{equation}
A_\theta(t)= \frac{\mu B_0 \theta_0 }{I\omega_0^2},\quad 0\le t< \tau_3  \, .
\end{equation}
At $t=\tau_3$, we flip the NV spin between $s=+1$ and $s=-1$ and then the equilibrium position $\bar{\theta}$ is changed
\begin{equation}
\delta\bar{\theta}(\tau_3)=\frac{2\mu B_c(\tau_3)\theta_0}{I\omega_0^2}\, .
\end{equation}
The rapid change of $\bar{\theta}$ kicks the oscillator, then the amplitude after the spin flipping satisfies
\begin{equation}
|A_\theta(0)-\delta\bar{\theta}(\tau_3)|\le A_\theta(t)\le |A_\theta(0)+\delta\bar{\theta}(\tau_3)|,\ t>\tau_3\, . 
\end{equation}
In our SGI model, we have $B_c(\tau_4)\sim B_c(\tau_3)\lesssim B_0$. Therefore, the bound of the oscillation amplitude can be estimated by $A_\theta(0)\lesssim A_\theta(t>\tau_3)\lesssim 3A_\theta(0)$.

At $t=\tau_4$, there is a mismatch of $\bar{\theta}(t)$ between the two SGI paths, namely $\delta\bar{\theta}(\tau_4)=2\mu B_c(\tau_4)\theta_0/I\omega_0^2$. The final mismatch $\delta\theta$ of $\theta(\tau_4)$ between the two paths can be roughly estimated by $\delta\bar{\theta}(\tau_4)$ and $A_\theta(\tau_4)$, namely
\begin{equation}\label{c9mismatchtheta}
\delta\theta<A_{\theta L}(\tau_4) + A_{\theta R}(\tau_4) + \delta\bar{\theta}(\tau_4) \lesssim \frac{8\mu}{I\omega_0^2}B_0\theta_0  \, ,
\end{equation}
where we have taken the upper bound of the oscillation amplitude $A_\theta(\tau_4)\lesssim 3\mu B_0\theta_0/I\omega_0^2$ for both the left and right paths.


\section{Quantum evolution of angular wave packets}\label{appC}

According to the canonical quantization procedure, we have $[\hat{\theta}, \hat{p}_\theta]=[\hat{\phi}, \hat{p}_\phi]=[\hat{\psi}, \hat{p}_\psi]=i\hbar$, and $[\hat{p}_\phi,\hat{p}_\psi]=[\hat{p}_\theta, \hat{p}_\phi]=[\hat{p}_\psi, \hat{p}_\theta]=0$, which are consistent with the commutation relations of angular momentum~\cite{barut1992magnetic}, i.e. $[\hat{L}_i,\hat{L}_j]=i\varepsilon_{ijk}\hat{L}_k$, where $\{i,j,k\}$ label $\{x,y,z\}$ or $\{1,2,3\}$. First, we define the shifted operators
\begin{equation}
\hat{\theta}' \equiv \hat{\theta}-\theta_0-\frac{\mu s B_c\theta_0}{I\omega_0^2}\, , \ \hat{p}'_\phi \equiv \hat{p}_\phi-\langle \hat{p}_\phi \rangle\, , \ \hat{p}'_\psi \equiv \hat{p}_\psi-\langle \hat{p}_\psi \rangle \, ,
\end{equation}
where $\langle \hat{p}_\phi \rangle=I\omega_0 {\rm cos}\theta_0$, $\langle \hat{p}_\psi \rangle=I\omega_0$ are the expectation values of $\hat{p}_\phi$ and $\hat{p}_\psi$.
The quantum Hamiltonian of rotational dynamics [Eq.(7) of the main text] can be written as
\begin{equation}\label{quantumhamiltonian}
\hat{H} \approx \frac{\hat{p}_\theta^2}{2I} +\frac{1}{2}I\omega^2_0\hat{\theta}'^2 + g(\hat{p}'_\phi, \hat{p}'_\psi, t) - f(\hat{p}'_\phi, \hat{p}'_\psi)\hat{\theta}' \, ,
\end{equation}
where the function $f$ and $g$ is defined as
\begin{align}
&g(p'_\phi, p'_\psi, t)=\frac{(p'_\phi-p'_\psi)^2}{2I \theta_0^2} - \omega_0 p'_\psi - \frac{\mu s B_c}{I\omega_0}(p'_\phi-p'_\psi)\, ,\nonumber\\
&f(p'_\phi, p'_\psi) = \frac{\omega_0}{\theta_0}(p'_\phi-p'_\psi) + \frac{(p'_\phi-p'_\psi)^2}{I\theta_0^3} \, .
\end{align}
We neglect higher order terms of $\hat{\theta}'$ based on assumption $\langle\hat{\theta}'\rangle \ll \theta_0\ll 1$ and we assume that the quantum fluctuation of momentum is small, namely $\Delta p_\phi \approx \Delta p_\psi\sim \hbar \ll I\omega_0$. 

Initially, the spin state is $s=0$, and the quantum wave packet of $\phi$ and $\psi$ degrees of freedom in the momentum space is set as Gaussian wave packets with width $\Delta p_\phi$ and $\Delta p_\psi$, respectively. The quantum state of the nutation mode $\theta$ is set as the ground state $\lvert0\rangle_{\bar{\theta}_0}$ in the harmonic trap with frequency $\omega_0$ and equilibrium position $\bar{\theta}_0$. So, the initial wave function of the angular d.o.f. is
\begin{equation}
\lvert \Psi(0) \rangle = \int \frac{d p'_\phi \, d p'_\psi}{\sqrt{2\pi \Delta p_\phi\Delta p_\psi}} \, e^{- \frac{(p'_\phi)^2}{4\Delta p_\phi^2}} e^{-\frac{(p'_\psi)^2}{4\Delta p_\psi^2}}  \lvert p'_\phi, p'_\psi \rangle \otimes |0\rangle_{\bar{\theta}_0} \, .
\end{equation}
From the Hamiltonian (\ref{quantumhamiltonian}), the initial equilibrium position $\bar{\theta}$ of the $\theta$ mode is
\begin{equation}
    \bar{\theta}_0(p'_\phi, p'_\psi)=\theta_0+\frac{f(p'_\phi, p'_\psi)}{I\omega_0^2}\, .
\end{equation}
We can see that the equilibrium position of the harmonic trap is also shifted by the quantum fluctuations $p'_\phi$ and $p'_\psi$. When the interferometer starts, the spin state is prepared as a superposition state of $s=+1$ and $s=-1$. Then $\bar{\theta}$ deviates from the initial equilibrium position $\bar{\theta}_0$ so that the ground state $|0\rangle_{\bar{\theta}_0}$ becomes a coherent state in the harmonic trap with new balance position
\begin{equation}
    \bar{\theta}(s,t,p'_\phi, p'_\psi)= \theta_0 + \frac{ \mu s B_c(t)\theta_0 + f(p'_\phi, p'_\psi) }{I\omega_0^2} \, .
\end{equation}
Therefore, the quantum evolution of wave packet is given by
\begin{align}\label{quantumstate}
&\lvert \Psi(t)\rangle = exp \left(-\frac{i}{\hbar} \int \hat{H} dt \right) \lvert \Psi(0) \rangle \nonumber \\
& = \int \frac{d p'_\phi \, d p'_\psi}{\sqrt{2\pi \Delta p_\phi\Delta p_\psi}} \, e^{- \frac{(p'_\phi)^2}{4\Delta p_\phi^2}} e^{-\frac{(p'_\psi)^2}{4\Delta p_\psi^2}} \nonumber\\
& \times exp \left[ -\frac{i}{\hbar}\int dt \, g(p'_\phi, p'_\psi, t) + f(p'_\phi, p'_\psi)\hat{\theta}' \right.  \nonumber\\
&+ \left. \frac{\hat{p}_\theta^2}{2I} +\frac{1}{2}I\omega^2_0\hat{\theta}'^2  \right]  \lvert p'_\phi, p'_\psi \rangle \otimes |\alpha\rangle \nonumber\\
&= \int \frac{d p'_\phi \, d p'_\psi}{\sqrt{2\pi \Delta p_\phi\Delta p_\psi}} \, e^{- \frac{(p'_\phi)^2}{4\Delta p_\phi^2}} e^{-\frac{(p'_\psi)^2}{4\Delta p_\psi^2}} \nonumber\\
& \times exp \left[ -\frac{i}{\hbar}\int dt \, g(p'_\phi, p'_\psi, t)\right]  \lvert p'_\phi, p'_\psi \rangle \otimes |\alpha(t)\rangle_{\bar{\theta}} \, ,
\end{align}
and, near $t=0$, the coherent state $\lvert \alpha(t)\rangle_{\bar{\theta}}$ is evolving as  
\begin{align}
\lvert \alpha(t)\rangle_{\bar{\theta}} &= exp \left[ \frac{-i}{\hbar}\int dt \, \frac{\hat{p}_\theta^2}{2I} +\frac{I\omega^2_0}{2}\hat{\theta}'^2 - f(p'_\phi, p'_\psi)\hat{\theta}' \right] |0\rangle_{\bar{\theta}_0} \nonumber\\
&= \bigg\vert  -\sqrt{\frac{I\omega_0}{2\hbar}} \frac{\mu s B_0\theta_0}{I\omega_0^2}  e^{-i\omega_0 t} \Bigg\rangle_{\bar{\theta}} \, .
\end{align}
Since both of $\bar{\theta}$ and $\bar{\theta}_0$ are shifted by $f/(I\omega^2_0)$, the quantum fluctuation of momentum $p_\phi$ and $p_\psi$ do not affect the initial amplitude of the nutation mode. The evolution of the amplitude $|\alpha|$ of the coherent state can be obtained from the analysis of classical amplitude estimation in Appendix \ref{appB}, where the final amplitude $|\alpha|$ satisfies
\begin{equation}\label{maximumalpha}
    |\alpha(\tau_4)|< \sqrt{\frac{I\omega_0}{2\hbar}}\frac{3\mu B_0 \theta_0}{I\omega_0^2}\, .
\end{equation}

For the quantum state of $\psi$ and $\phi$, since the Hamiltonian consists only of quadratic terms of $\hat{p}'_\phi$ and $\hat{p}'_\psi$, the wave packet of $\phi$ and $\psi$ remains in Gaussian form during evolution. And because of $[\hat{p}'_\phi,\hat{H}]=0$ and $[\hat{p}'_\psi,\hat{H}]=0$, the momentum uncertainties $\Delta p_\phi$ and $\Delta p_\psi$ are invariant. 

\section{Spin contrast computation}\label{appD}

From the solution (\ref{quantumstate}) of the angular quantum state, we can figure out the SGI contrast for the rotational dynamics, which is given by
\begin{align}\label{contrastquantum}
C &= |\langle \Psi_L(\tau_4)|\Psi_R(\tau_4)\rangle| = \int \frac{d p'_\phi \, d p'_\psi d p''_\phi \, d p''_\psi}{2\pi \Delta p_\phi\Delta p_\psi}  \nonumber \\
& \times e^{- \frac{(p'_\phi)^2+(p''_\phi)^2}{4\Delta p_\phi^2}} e^{-\frac{(p'_\psi+(p''_\psi)^2)^2}{4\Delta p_\psi^2}} \nonumber \\
& \times exp\left[ -\frac{i}{\hbar} \int_0^{\tau_4} dt (g_L(p'_\phi, p'_\psi, t)-g_R(p''_\phi, p''_\psi, t))  \right] \nonumber\\
& \times \langle p'_\phi, p'_\psi \lvert p''_\phi, p''_\psi \rangle \times \langle \alpha_L(p'_\phi, p'_\psi,\tau_4) \lvert \alpha_R(p''_\phi, p''_\psi, \tau_4) \rangle \nonumber\\
&=\int \frac{d p'_\phi \, d p'_\psi }{2\pi \Delta p_\phi\Delta p_\psi} \, e^{- \frac{(p'_\phi)^2}{2\Delta p_\phi^2}} e^{-\frac{(p'_\psi)^2}{2\Delta p_\psi^2}}  \nonumber \\ 
&\times exp\left[ -\frac{i}{\hbar} \int_0^{\tau_4} dt (g_L(p'_\phi, p'_\psi, t)-g_R(p'_\phi, p'_\psi, t))  \right] \nonumber\\
&\times |\leftindex_{\bar{\theta}_L}{\langle} \alpha_L(p'_\phi, p'_\psi,\tau_4) \lvert \alpha_R(p'_\phi, p'_\psi, \tau_4) \rangle_{\bar{\theta}_R}| \, ,
\end{align}
in which we have
\begin{align}\label{b8}
&\int_0^{\tau_4} dt\,[ g_L(p'_\phi, p'_\psi, t)-g_R(p'_\phi, p'_\psi, t)] \nonumber\\
&=\int_0^{\tau_4} dt \left( \frac{\mu s B_c}{I\omega_0}\bigg|_L - \frac{\mu s B_c}{I\omega_0}\bigg|_R \right) (p'_\phi-p'_\psi) \nonumber\\
&= \delta \phi \times (p'_\phi-p'_\psi) \, .
\end{align} 
$\delta\phi$, evaluated in Eq.(15) of the main text, is the mismatch between the classical trajectory of $\phi$ of the left and right SGI arms. The module of the inner product $\leftindex_{\bar{\theta}_L}{\langle} \alpha_L(p'_\phi, p'_\psi,\tau_4) \lvert \alpha_R(p'_\phi, p'_\psi, \tau_4) \rangle_{\bar{\theta}_R}$ of two coherent states is determined by the distance of the complex variables $\alpha_L$ and $\alpha_R$, namely, 
\begin{equation}\label{b9}
|\leftindex_{\bar{\theta}_L}{\langle} \alpha_L \lvert \alpha_R \rangle_{\bar{\theta}_R}| 
 = exp\left[ -\frac{1}{2} |\alpha_L-(\alpha_R +\delta X)|^2 \right]\, ,
\end{equation}
where 
\begin{equation}\label{deltaX}
\delta X\equiv \sqrt{\frac{I\omega_0}{2\hbar}} |\bar{\theta}_L-\bar{\theta}_R| = \sqrt{\frac{I\omega_0}{2\hbar}} \frac{2\mu B_c(\tau_4)\theta_0}{I\omega_0^2} \, .
\end{equation}
Notably, the equilibrium position deviations due to the quantum fluctuation $p'_\phi$ and $p'_\psi$ are the same for left and right SGI paths, thus the overlap of the two coherent state $|\leftindex_{\bar{\theta}_L}{\langle} \alpha_L \lvert \alpha_R \rangle_{\bar{\theta}_R}|$ is independent on the variable $p'_\phi$ and $p'_\psi$.
\begin{figure}[ht]
    \centering
    \includegraphics[width=0.8\linewidth]{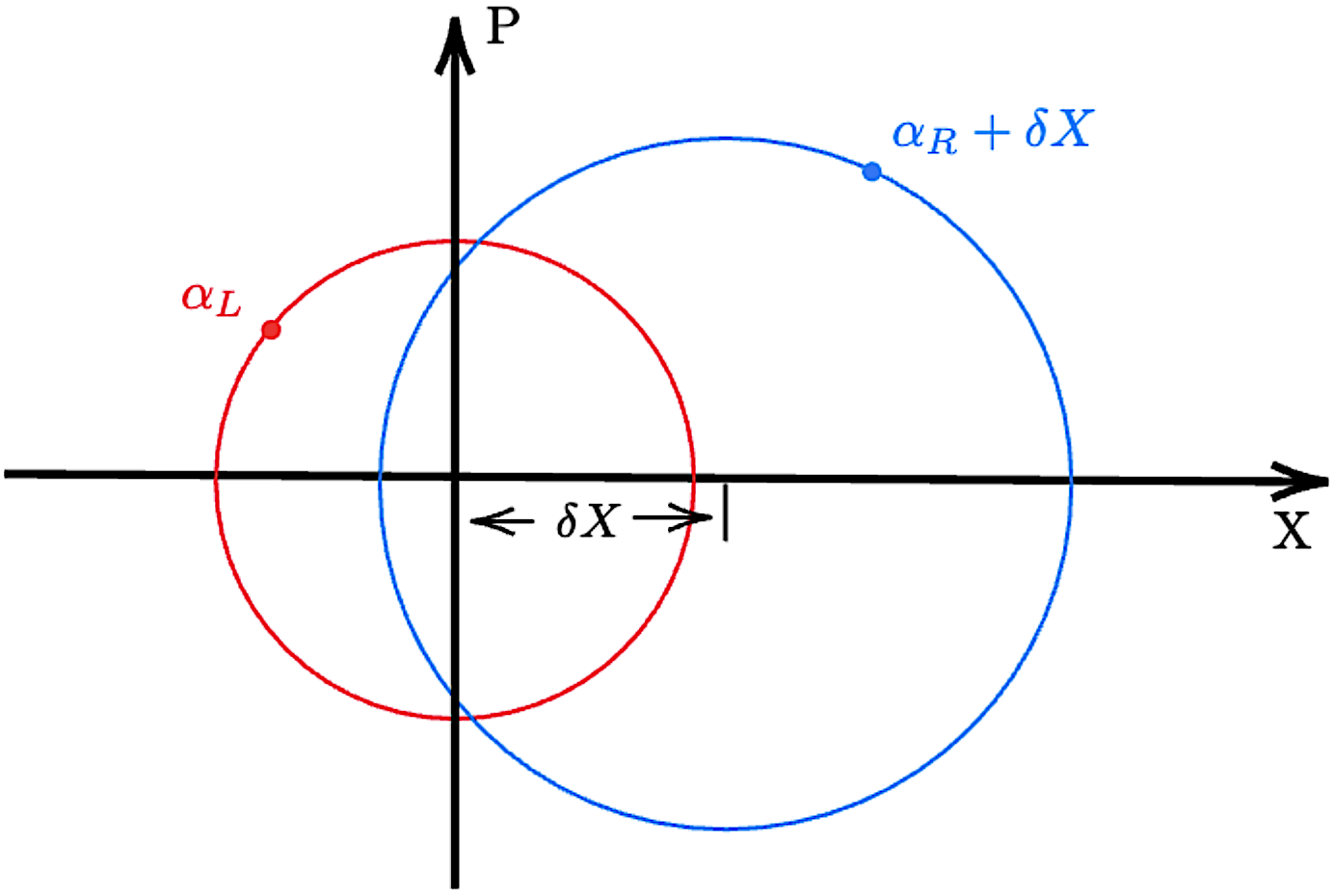}
    \caption{Sketch of coherent state $|\alpha_L\rangle$ and $|\alpha_R\rangle$ in the phase space $\{X=\sqrt{I\omega_0/(2\hbar)}\theta, P=\sqrt{1/(I\omega_02\hbar)}\, p_\theta\}$ at $t\approx \tau_4$. The red and blue circle represents the evolution of the two coherent states. The  mismatch of dimensionless equilibrium position $\delta X=\sqrt{I\omega_0/(2\hbar)}\,\delta\bar{\theta}(\tau_4)$.}
    \label{coherentstate}
\end{figure}

Plugging (\ref{b8}) and (\ref{b9}) into the contrast formula (\ref{contrastquantum}), we end up with the contrast
\begin{align}\label{contrastappendix}
&C = |\langle \alpha_L \lvert \alpha_R \rangle| \int \frac{d p'_\phi \, d p'_\psi }{2\pi \Delta p_\phi\Delta p_\psi} \, e^{- \frac{(p'_\phi)^2}{2\Delta p_\phi^2}-\frac{(p'_\psi)^2}{2\Delta p_\psi^2}+\frac{i}{\hbar}\delta\phi (p'_\phi-p'_\psi)}  \nonumber \\ 
&=exp\left[-\frac{1}{2} \left( \delta\phi^2 \frac{\Delta p_\phi^2}{\hbar^2} +\delta\psi^2 \frac{\Delta p_\psi^2}{\hbar^2} \right) \right] |\langle \alpha_L \lvert \alpha_R \rangle| \, .
\end{align}

Using (\ref{maximumalpha}, \ref{b9}, \ref{deltaX}), we obtain the lower bound of the overlap bwteen the two coherent state
\begin{align}\label{alphalowerbound}
    &|\leftindex_{\bar{\theta}_L}{\langle} \alpha_L \lvert \alpha_R \rangle_{\bar{\theta}_R}|  > exp \left[ -\frac{1}{2}\left(|\alpha_L|+|\alpha_R|+\delta X\right)^2 \right]  \nonumber\\
    &> exp\left[ -\frac{1}{2}\frac{I\omega_0}{2\hbar} \left(\frac{8\mu B_0 \theta_0}{ I\omega_0^2 } \right)^2 \right]
\end{align}
Therefore, the lower bound of the contrast is
\begin{equation}\label{contrastbound}
C >  e^{-\frac{1}{2}\delta\phi^2 \frac{\Delta p_\phi^2}{\hbar^2}} e^{-\frac{1}{2}\delta\psi^2 \frac{\Delta p_\psi^2}{\hbar^2}}  exp \left( -\frac{16\mu^2B_0^2\theta_0^2}{\hbar I \omega_0^3} \right) \, .
\end{equation}

\section{Finite temperature correction}\label{appE}

The contrast (\ref{contrastbound}) is calculated in the zero-temperature case, that the initial state of nutation mode is the ground state of the harmonic trap. Considering a finite temperature $T$, the initial state of nutation mode can be described by the density matrix operator in the coherent state basis
\begin{equation}
    \hat{\rho}_{th} = \int \frac{d^2\alpha}{\pi} \frac{e^{-\frac{|\alpha|^2}{n}}}{n}|\alpha\rangle\langle\alpha|\, ,
\end{equation}
in which $n\equiv k_B T/(\hbar\omega_0)$ is the occupation number. Here we take the definition of the final overlap between two thermally coherent states as $tr[ \hat{D}(\zeta_L) \hat{\rho}_{th} \hat{D}^\dagger(\zeta_R)]$, where $\hat{D}(\zeta)$ is the displacement operator that satisfies $\hat{D}(\zeta_L)|0\rangle_{\bar{\theta}_0}=|\alpha_L\rangle_{\bar{\theta}}$ and $\hat{D}(\zeta_R)|0\rangle_{\bar{\theta}_0}=|\alpha_R\rangle_{\bar{\theta}}$. Then, we have\cite{steiner2024pentacene}
\begin{equation}
    tr\left[ \hat{D}(\zeta_L) \hat{\rho}_{th} \hat{D}^\dagger(\zeta_R) \right]= exp\left[\varphi-|\Delta\zeta|^2\left(\frac{1}{2}+n\right)\right]\, ,
\end{equation}
where the phase factor $\varphi\equiv (\zeta_L\zeta_R^*-\zeta_L^*\zeta_R)/2$ is a imaginary number so it does not contribute on the contrast. Therefore, considering thermal effect, the equation (\ref{b9}) and (\ref{alphalowerbound}) should be rewrite as
\begin{align}
|\langle \alpha_L \lvert \alpha_R \rangle|_{th} 
 &= exp\left[ -\left(\frac{1}{2}+n\right) |\alpha_L-(\alpha_R +\delta X)|^2 \right]\nonumber\\
 & > exp \left[ -\left(\frac{1}{2}+n\right)\frac{32\mu^2B_0^2\theta_0^2}{\hbar I \omega_0^3}\right]\, .
\end{align}
The lower bound of the spin contrast in finite temperature case is given by
\begin{equation}\label{contrastthermalbound}
C_{th} >  exp \left[ -\frac{\delta\phi^2\Delta p_\phi^2}{2\hbar^2} - \frac{\delta\psi^2\Delta p_\psi^2}{2\hbar^2} - \frac{16(1+2n)\mu^2B_0^2\theta_0^2}{\hbar I \omega_0^3} \right] \, .
\end{equation}


\bibliography{reference}

\begin{thebibliography}{66}%
\makeatletter
\providecommand \@ifxundefined [1]{%
 \@ifx{#1\undefined}
}%
\providecommand \@ifnum [1]{%
 \ifnum #1\expandafter \@firstoftwo
 \else \expandafter \@secondoftwo
 \fi
}%
\providecommand \@ifx [1]{%
 \ifx #1\expandafter \@firstoftwo
 \else \expandafter \@secondoftwo
 \fi
}%
\providecommand \natexlab [1]{#1}%
\providecommand \enquote  [1]{``#1''}%
\providecommand \bibnamefont  [1]{#1}%
\providecommand \bibfnamefont [1]{#1}%
\providecommand \citenamefont [1]{#1}%
\providecommand \href@noop [0]{\@secondoftwo}%
\providecommand \href [0]{\begingroup \@sanitize@url \@href}%
\providecommand \@href[1]{\@@startlink{#1}\@@href}%
\providecommand \@@href[1]{\endgroup#1\@@endlink}%
\providecommand \@sanitize@url [0]{\catcode `\\12\catcode `\$12\catcode
  `\&12\catcode `\#12\catcode `\^12\catcode `\_12\catcode `\%12\relax}%
\providecommand \@@startlink[1]{}%
\providecommand \@@endlink[0]{}%
\providecommand \url  [0]{\begingroup\@sanitize@url \@url }%
\providecommand \@url [1]{\endgroup\@href {#1}{\urlprefix }}%
\providecommand \urlprefix  [0]{URL }%
\providecommand \Eprint [0]{\href }%
\providecommand \doibase [0]{https://doi.org/}%
\providecommand \selectlanguage [0]{\@gobble}%
\providecommand \bibinfo  [0]{\@secondoftwo}%
\providecommand \bibfield  [0]{\@secondoftwo}%
\providecommand \translation [1]{[#1]}%
\providecommand \BibitemOpen [0]{}%
\providecommand \bibitemStop [0]{}%
\providecommand \bibitemNoStop [0]{.\EOS\space}%
\providecommand \EOS [0]{\spacefactor3000\relax}%
\providecommand \BibitemShut  [1]{\csname bibitem#1\endcsname}%
\let\auto@bib@innerbib\@empty
\bibitem [{\citenamefont {Margalit}\ \emph {et~al.}(2021)\citenamefont
  {Margalit}, \citenamefont {Dobkowski}, \citenamefont {Zhou}, \citenamefont
  {Amit}, \citenamefont {Japha}, \citenamefont {Moukouri}, \citenamefont
  {Rohrlich}, \citenamefont {Mazumdar}, \citenamefont {Bose}, \citenamefont
  {Henkel},\ and\ \citenamefont {Folman}}]{SGI_experiment}%
  \BibitemOpen
  \bibfield  {author} {\bibinfo {author} {\bibfnamefont {Y.}~\bibnamefont
  {Margalit}}, \bibinfo {author} {\bibfnamefont {O.}~\bibnamefont {Dobkowski}},
  \bibinfo {author} {\bibfnamefont {Z.}~\bibnamefont {Zhou}}, \bibinfo {author}
  {\bibfnamefont {O.}~\bibnamefont {Amit}}, \bibinfo {author} {\bibfnamefont
  {Y.}~\bibnamefont {Japha}}, \bibinfo {author} {\bibfnamefont
  {S.}~\bibnamefont {Moukouri}}, \bibinfo {author} {\bibfnamefont
  {D.}~\bibnamefont {Rohrlich}}, \bibinfo {author} {\bibfnamefont
  {A.}~\bibnamefont {Mazumdar}}, \bibinfo {author} {\bibfnamefont
  {S.}~\bibnamefont {Bose}}, \bibinfo {author} {\bibfnamefont {C.}~\bibnamefont
  {Henkel}},\ and\ \bibinfo {author} {\bibfnamefont {R.}~\bibnamefont
  {Folman}},\ }\bibfield  {title} {\bibinfo {title} {Realization of a complete
  stern-gerlach interferometer: Toward a test of quantum gravity},\ }\href
  {https://doi.org/10.1126/sciadv.abg2879} {\bibfield  {journal} {\bibinfo
  {journal} {Science Advances}\ }\textbf {\bibinfo {volume} {7}},\ \bibinfo
  {pages} {eabg2879} (\bibinfo {year} {2021})}\BibitemShut {NoStop}%
\bibitem [{\citenamefont {Amit}\ \emph {et~al.}(2019)\citenamefont {Amit},
  \citenamefont {Margalit}, \citenamefont {Dobkowski}, \citenamefont {Zhou},
  \citenamefont {Japha}, \citenamefont {Zimmermann}, \citenamefont {Efremov},
  \citenamefont {Narducci}, \citenamefont {Rasel}, \citenamefont {Schleich}
  \emph {et~al.}}]{amit2019t}%
  \BibitemOpen
  \bibfield  {author} {\bibinfo {author} {\bibfnamefont {O.}~\bibnamefont
  {Amit}}, \bibinfo {author} {\bibfnamefont {Y.}~\bibnamefont {Margalit}},
  \bibinfo {author} {\bibfnamefont {O.}~\bibnamefont {Dobkowski}}, \bibinfo
  {author} {\bibfnamefont {Z.}~\bibnamefont {Zhou}}, \bibinfo {author}
  {\bibfnamefont {Y.}~\bibnamefont {Japha}}, \bibinfo {author} {\bibfnamefont
  {M.}~\bibnamefont {Zimmermann}}, \bibinfo {author} {\bibfnamefont {M.~A.}\
  \bibnamefont {Efremov}}, \bibinfo {author} {\bibfnamefont {F.~A.}\
  \bibnamefont {Narducci}}, \bibinfo {author} {\bibfnamefont {E.~M.}\
  \bibnamefont {Rasel}}, \bibinfo {author} {\bibfnamefont {W.~P.}\ \bibnamefont
  {Schleich}}, \emph {et~al.},\ }\bibfield  {title} {\bibinfo {title} {T 3
  stern-gerlach matter-wave interferometer},\ }\href
  {https://doi.org/10.1103/PhysRevLett.123.083601} {\bibfield  {journal}
  {\bibinfo  {journal} {Physical review letters}\ }\textbf {\bibinfo {volume}
  {123}},\ \bibinfo {pages} {083601} (\bibinfo {year} {2019})}\BibitemShut
  {NoStop}%
\bibitem [{\citenamefont {Wan}\ \emph {et~al.}(2016)\citenamefont {Wan},
  \citenamefont {Scala}, \citenamefont {Bose}, \citenamefont {Frangeskou},
  \citenamefont {Rahman}, \citenamefont {Morley}, \citenamefont {Barker},\ and\
  \citenamefont {Kim}}]{WanPRA16_GM}%
  \BibitemOpen
  \bibfield  {author} {\bibinfo {author} {\bibfnamefont {C.}~\bibnamefont
  {Wan}}, \bibinfo {author} {\bibfnamefont {M.}~\bibnamefont {Scala}}, \bibinfo
  {author} {\bibfnamefont {S.}~\bibnamefont {Bose}}, \bibinfo {author}
  {\bibfnamefont {A.~C.}\ \bibnamefont {Frangeskou}}, \bibinfo {author}
  {\bibfnamefont {A.~T. M.~A.}\ \bibnamefont {Rahman}}, \bibinfo {author}
  {\bibfnamefont {G.~W.}\ \bibnamefont {Morley}}, \bibinfo {author}
  {\bibfnamefont {P.~F.}\ \bibnamefont {Barker}},\ and\ \bibinfo {author}
  {\bibfnamefont {M.~S.}\ \bibnamefont {Kim}},\ }\bibfield  {title} {\bibinfo
  {title} {Tolerance in the {R}amsey interference of a trapped nanodiamond},\
  }\href
  {http://0-link.aps.org.pugwash.lib.warwick.ac.uk/doi/10.1103/PhysRevA.93.043852}
  {\bibfield  {journal} {\bibinfo  {journal} {Phys. Rev. A}\ }\textbf {\bibinfo
  {volume} {93}},\ \bibinfo {pages} {043852} (\bibinfo {year}
  {2016})}\BibitemShut {NoStop}%
\bibitem [{\citenamefont {Scala}\ \emph {et~al.}(2013)\citenamefont {Scala},
  \citenamefont {Kim}, \citenamefont {Morley}, \citenamefont {Barker},\ and\
  \citenamefont {Bose}}]{Scala13_GM}%
  \BibitemOpen
  \bibfield  {author} {\bibinfo {author} {\bibfnamefont {M.}~\bibnamefont
  {Scala}}, \bibinfo {author} {\bibfnamefont {M.~S.}\ \bibnamefont {Kim}},
  \bibinfo {author} {\bibfnamefont {G.~W.}\ \bibnamefont {Morley}}, \bibinfo
  {author} {\bibfnamefont {P.~F.}\ \bibnamefont {Barker}},\ and\ \bibinfo
  {author} {\bibfnamefont {S.}~\bibnamefont {Bose}},\ }\bibfield  {title}
  {\bibinfo {title} {Matter-wave interferometry of a levitated thermal
  nano-oscillator induced and probed by a spin},\ }\href
  {http://link.aps.org/doi/10.1103/PhysRevLett.111.180403} {\bibfield
  {journal} {\bibinfo  {journal} {Phys. Rev. Lett.}\ }\textbf {\bibinfo
  {volume} {111}},\ \bibinfo {pages} {180403} (\bibinfo {year}
  {2013})}\BibitemShut {NoStop}%
\bibitem [{\citenamefont {Pedernales}\ \emph {et~al.}(2020)\citenamefont
  {Pedernales}, \citenamefont {Morley},\ and\ \citenamefont
  {Plenio}}]{Pedernales:2020nmf}%
  \BibitemOpen
  \bibfield  {author} {\bibinfo {author} {\bibfnamefont {J.~S.}\ \bibnamefont
  {Pedernales}}, \bibinfo {author} {\bibfnamefont {G.~W.}\ \bibnamefont
  {Morley}},\ and\ \bibinfo {author} {\bibfnamefont {M.~B.}\ \bibnamefont
  {Plenio}},\ }\bibfield  {title} {\bibinfo {title} {{Motional Dynamical
  Decoupling for Interferometry with Macroscopic Particles}},\ }\href
  {https://doi.org/10.1103/PhysRevLett.125.023602} {\bibfield  {journal}
  {\bibinfo  {journal} {Phys. Rev. Lett.}\ }\textbf {\bibinfo {volume} {125}},\
  \bibinfo {pages} {023602} (\bibinfo {year} {2020})}\BibitemShut {NoStop}%
\bibitem [{\citenamefont {Marshman}\ \emph {et~al.}(2022)\citenamefont
  {Marshman}, \citenamefont {Mazumdar}, \citenamefont {Folman},\ and\
  \citenamefont {Bose}}]{Marshman:2021wyk}%
  \BibitemOpen
  \bibfield  {author} {\bibinfo {author} {\bibfnamefont {R.~J.}\ \bibnamefont
  {Marshman}}, \bibinfo {author} {\bibfnamefont {A.}~\bibnamefont {Mazumdar}},
  \bibinfo {author} {\bibfnamefont {R.}~\bibnamefont {Folman}},\ and\ \bibinfo
  {author} {\bibfnamefont {S.}~\bibnamefont {Bose}},\ }\bibfield  {title}
  {\bibinfo {title} {{Constructing nano-object quantum superpositions with a
  Stern-Gerlach interferometer}},\ }\href
  {https://doi.org/10.1103/PhysRevResearch.4.023087} {\bibfield  {journal}
  {\bibinfo  {journal} {Phys. Rev. Res.}\ }\textbf {\bibinfo {volume} {4}},\
  \bibinfo {pages} {023087} (\bibinfo {year} {2022})},\ \Eprint
  {https://arxiv.org/abs/2105.01094} {arXiv:2105.01094 [quant-ph]} \BibitemShut
  {NoStop}%
\bibitem [{\citenamefont {Wu}\ \emph {et~al.}(2023)\citenamefont {Wu},
  \citenamefont {Toro\v{s}}, \citenamefont {Bose},\ and\ \citenamefont
  {Mazumdar}}]{Wu:2022rdv}%
  \BibitemOpen
  \bibfield  {author} {\bibinfo {author} {\bibfnamefont {M.-Z.}\ \bibnamefont
  {Wu}}, \bibinfo {author} {\bibfnamefont {M.}~\bibnamefont {Toro\v{s}}},
  \bibinfo {author} {\bibfnamefont {S.}~\bibnamefont {Bose}},\ and\ \bibinfo
  {author} {\bibfnamefont {A.}~\bibnamefont {Mazumdar}},\ }\bibfield  {title}
  {\bibinfo {title} {{Quantum gravitational sensor for space debris}},\ }\href
  {https://doi.org/10.1103/PhysRevD.107.104053} {\bibfield  {journal} {\bibinfo
   {journal} {Phys. Rev. D}\ }\textbf {\bibinfo {volume} {107}},\ \bibinfo
  {pages} {104053} (\bibinfo {year} {2023})},\ \Eprint
  {https://arxiv.org/abs/2211.15695} {arXiv:2211.15695 [gr-qc]} \BibitemShut
  {NoStop}%
\bibitem [{\citenamefont {Barker}\ \emph {et~al.}(2022)\citenamefont {Barker},
  \citenamefont {Bose}, \citenamefont {Marshman},\ and\ \citenamefont
  {Mazumdar}}]{Barker:2022mdz}%
  \BibitemOpen
  \bibfield  {author} {\bibinfo {author} {\bibfnamefont {P.~F.}\ \bibnamefont
  {Barker}}, \bibinfo {author} {\bibfnamefont {S.}~\bibnamefont {Bose}},
  \bibinfo {author} {\bibfnamefont {R.~J.}\ \bibnamefont {Marshman}},\ and\
  \bibinfo {author} {\bibfnamefont {A.}~\bibnamefont {Mazumdar}},\ }\bibfield
  {title} {\bibinfo {title} {{Entanglement based tomography to probe new
  macroscopic forces}},\ }\href {https://doi.org/10.1103/PhysRevD.106.L041901}
  {\bibfield  {journal} {\bibinfo  {journal} {Phys. Rev. D}\ }\textbf {\bibinfo
  {volume} {106}},\ \bibinfo {pages} {L041901} (\bibinfo {year} {2022})},\
  \Eprint {https://arxiv.org/abs/2203.00038} {arXiv:2203.00038 [hep-ph]}
  \BibitemShut {NoStop}%
\bibitem [{\citenamefont {Marshman}\ \emph
  {et~al.}(2020{\natexlab{a}})\citenamefont {Marshman}, \citenamefont
  {Mazumdar}, \citenamefont {Morley}, \citenamefont {Barker}, \citenamefont
  {Hoekstra},\ and\ \citenamefont {Bose}}]{Marshman:2018upe}%
  \BibitemOpen
  \bibfield  {author} {\bibinfo {author} {\bibfnamefont {R.~J.}\ \bibnamefont
  {Marshman}}, \bibinfo {author} {\bibfnamefont {A.}~\bibnamefont {Mazumdar}},
  \bibinfo {author} {\bibfnamefont {G.~W.}\ \bibnamefont {Morley}}, \bibinfo
  {author} {\bibfnamefont {P.~F.}\ \bibnamefont {Barker}}, \bibinfo {author}
  {\bibfnamefont {S.}~\bibnamefont {Hoekstra}},\ and\ \bibinfo {author}
  {\bibfnamefont {S.}~\bibnamefont {Bose}},\ }\bibfield  {title} {\bibinfo
  {title} {{Mesoscopic Interference for Metric and Curvature (MIMAC) $\&$
  Gravitational Wave Detection}},\ }\href
  {https://doi.org/10.1088/1367-2630/ab9f6c} {\bibfield  {journal} {\bibinfo
  {journal} {New J. Phys.}\ }\textbf {\bibinfo {volume} {22}},\ \bibinfo
  {pages} {083012} (\bibinfo {year} {2020}{\natexlab{a}})},\ \Eprint
  {https://arxiv.org/abs/1807.10830} {arXiv:1807.10830 [gr-qc]} \BibitemShut
  {NoStop}%
\bibitem [{\citenamefont {Bose}\ \emph {et~al.}(2017)\citenamefont {Bose},
  \citenamefont {Mazumdar}, \citenamefont {Morley}, \citenamefont {Ulbricht},
  \citenamefont {Toro\v{s}}, \citenamefont {Paternostro}, \citenamefont
  {Geraci}, \citenamefont {Barker}, \citenamefont {Kim},\ and\ \citenamefont
  {Milburn}}]{Bose:2017nin}%
  \BibitemOpen
  \bibfield  {author} {\bibinfo {author} {\bibfnamefont {S.}~\bibnamefont
  {Bose}}, \bibinfo {author} {\bibfnamefont {A.}~\bibnamefont {Mazumdar}},
  \bibinfo {author} {\bibfnamefont {G.~W.}\ \bibnamefont {Morley}}, \bibinfo
  {author} {\bibfnamefont {H.}~\bibnamefont {Ulbricht}}, \bibinfo {author}
  {\bibfnamefont {M.}~\bibnamefont {Toro\v{s}}}, \bibinfo {author}
  {\bibfnamefont {M.}~\bibnamefont {Paternostro}}, \bibinfo {author}
  {\bibfnamefont {A.}~\bibnamefont {Geraci}}, \bibinfo {author} {\bibfnamefont
  {P.}~\bibnamefont {Barker}}, \bibinfo {author} {\bibfnamefont {M.~S.}\
  \bibnamefont {Kim}},\ and\ \bibinfo {author} {\bibfnamefont {G.}~\bibnamefont
  {Milburn}},\ }\bibfield  {title} {\bibinfo {title} {{Spin Entanglement
  Witness for Quantum Gravity}},\ }\href
  {https://doi.org/10.1103/PhysRevLett.119.240401} {\bibfield  {journal}
  {\bibinfo  {journal} {Phys. Rev. Lett.}\ }\textbf {\bibinfo {volume} {119}},\
  \bibinfo {pages} {240401} (\bibinfo {year} {2017})},\ \Eprint
  {https://arxiv.org/abs/1707.06050} {arXiv:1707.06050 [quant-ph]} \BibitemShut
  {NoStop}%
\bibitem [{\citenamefont {Marshman}\ \emph
  {et~al.}(2020{\natexlab{b}})\citenamefont {Marshman}, \citenamefont
  {Mazumdar},\ and\ \citenamefont {Bose}}]{marshman2020locality}%
  \BibitemOpen
  \bibfield  {author} {\bibinfo {author} {\bibfnamefont {R.~J.}\ \bibnamefont
  {Marshman}}, \bibinfo {author} {\bibfnamefont {A.}~\bibnamefont {Mazumdar}},\
  and\ \bibinfo {author} {\bibfnamefont {S.}~\bibnamefont {Bose}},\ }\bibfield
  {title} {\bibinfo {title} {Locality and entanglement in table-top testing of
  the quantum nature of linearized gravity},\ }\href
  {https://doi.org/10.1103/PhysRevA.101.052110} {\bibfield  {journal} {\bibinfo
   {journal} {Physical Review A}\ }\textbf {\bibinfo {volume} {101}},\ \bibinfo
  {pages} {052110} (\bibinfo {year} {2020}{\natexlab{b}})}\BibitemShut
  {NoStop}%
\bibitem [{\citenamefont {Bose}\ \emph {et~al.}(2023)\citenamefont {Bose},
  \citenamefont {Mazumdar}, \citenamefont {Schut},\ and\ \citenamefont
  {Toro{\v{s}}}}]{bose2023entanglement}%
  \BibitemOpen
  \bibfield  {author} {\bibinfo {author} {\bibfnamefont {S.}~\bibnamefont
  {Bose}}, \bibinfo {author} {\bibfnamefont {A.}~\bibnamefont {Mazumdar}},
  \bibinfo {author} {\bibfnamefont {M.}~\bibnamefont {Schut}},\ and\ \bibinfo
  {author} {\bibfnamefont {M.}~\bibnamefont {Toro{\v{s}}}},\ }\bibfield
  {title} {\bibinfo {title} {Entanglement witness for the weak equivalence
  principle},\ }\href {https://doi.org/10.3390/e25030448} {\bibfield  {journal}
  {\bibinfo  {journal} {Entropy}\ }\textbf {\bibinfo {volume} {25}},\ \bibinfo
  {pages} {448} (\bibinfo {year} {2023})}\BibitemShut {NoStop}%
\bibitem [{\citenamefont {Carney}\ \emph {et~al.}(2019)\citenamefont {Carney},
  \citenamefont {Stamp},\ and\ \citenamefont {Taylor}}]{Carney_2019}%
  \BibitemOpen
  \bibfield  {author} {\bibinfo {author} {\bibfnamefont {D.}~\bibnamefont
  {Carney}}, \bibinfo {author} {\bibfnamefont {P.~C.~E.}\ \bibnamefont
  {Stamp}},\ and\ \bibinfo {author} {\bibfnamefont {J.~M.}\ \bibnamefont
  {Taylor}},\ }\bibfield  {title} {\bibinfo {title} {Tabletop experiments for
  quantum gravity: a user's manual},\ }\href
  {https://doi.org/10.1088/1361-6382/aaf9ca} {\bibfield  {journal} {\bibinfo
  {journal} {Class. Quant. Grav.}\ }\textbf {\bibinfo {volume} {36}},\ \bibinfo
  {pages} {034001} (\bibinfo {year} {2019})}\BibitemShut {NoStop}%
\bibitem [{\citenamefont {Carney}(2022)}]{Carney:2021vvt}%
  \BibitemOpen
  \bibfield  {author} {\bibinfo {author} {\bibfnamefont {D.}~\bibnamefont
  {Carney}},\ }\bibfield  {title} {\bibinfo {title} {{Newton, entanglement, and
  the graviton}},\ }\href {https://doi.org/10.1103/PhysRevD.105.024029}
  {\bibfield  {journal} {\bibinfo  {journal} {Phys. Rev. D}\ }\textbf {\bibinfo
  {volume} {105}},\ \bibinfo {pages} {024029} (\bibinfo {year} {2022})},\
  \Eprint {https://arxiv.org/abs/2108.06320} {arXiv:2108.06320 [quant-ph]}
  \BibitemShut {NoStop}%
\bibitem [{\citenamefont {Danielson}\ \emph {et~al.}(2022)\citenamefont
  {Danielson}, \citenamefont {Satishchandran},\ and\ \citenamefont
  {Wald}}]{Danielson:2021egj}%
  \BibitemOpen
  \bibfield  {author} {\bibinfo {author} {\bibfnamefont {D.~L.}\ \bibnamefont
  {Danielson}}, \bibinfo {author} {\bibfnamefont {G.}~\bibnamefont
  {Satishchandran}},\ and\ \bibinfo {author} {\bibfnamefont {R.~M.}\
  \bibnamefont {Wald}},\ }\bibfield  {title} {\bibinfo {title}
  {{Gravitationally mediated entanglement: Newtonian field versus gravitons}},\
  }\href {https://doi.org/10.1103/PhysRevD.105.086001} {\bibfield  {journal}
  {\bibinfo  {journal} {Phys. Rev. D}\ }\textbf {\bibinfo {volume} {105}},\
  \bibinfo {pages} {086001} (\bibinfo {year} {2022})}\BibitemShut {NoStop}%
\bibitem [{\citenamefont {Christodoulou}\ \emph {et~al.}(2023)\citenamefont
  {Christodoulou}, \citenamefont {Di~Biagio}, \citenamefont {Aspelmeyer},
  \citenamefont {Brukner}, \citenamefont {Rovelli},\ and\ \citenamefont
  {Howl}}]{christodoulou2023locally}%
  \BibitemOpen
  \bibfield  {author} {\bibinfo {author} {\bibfnamefont {M.}~\bibnamefont
  {Christodoulou}}, \bibinfo {author} {\bibfnamefont {A.}~\bibnamefont
  {Di~Biagio}}, \bibinfo {author} {\bibfnamefont {M.}~\bibnamefont
  {Aspelmeyer}}, \bibinfo {author} {\bibfnamefont {{\v{C}}.}~\bibnamefont
  {Brukner}}, \bibinfo {author} {\bibfnamefont {C.}~\bibnamefont {Rovelli}},\
  and\ \bibinfo {author} {\bibfnamefont {R.}~\bibnamefont {Howl}},\ }\bibfield
  {title} {\bibinfo {title} {Locally mediated entanglement in linearized
  quantum gravity},\ }\href {https://doi.org/10.1103/PhysRevLett.130.100202}
  {\bibfield  {journal} {\bibinfo  {journal} {Physical Review Letters}\
  }\textbf {\bibinfo {volume} {130}},\ \bibinfo {pages} {100202} (\bibinfo
  {year} {2023})}\BibitemShut {NoStop}%
\bibitem [{\citenamefont {Schut}\ \emph {et~al.}(2022)\citenamefont {Schut},
  \citenamefont {Tilly}, \citenamefont {Marshman}, \citenamefont {Bose},\ and\
  \citenamefont {Mazumdar}}]{Schut:2021svd}%
  \BibitemOpen
  \bibfield  {author} {\bibinfo {author} {\bibfnamefont {M.}~\bibnamefont
  {Schut}}, \bibinfo {author} {\bibfnamefont {J.}~\bibnamefont {Tilly}},
  \bibinfo {author} {\bibfnamefont {R.~J.}\ \bibnamefont {Marshman}}, \bibinfo
  {author} {\bibfnamefont {S.}~\bibnamefont {Bose}},\ and\ \bibinfo {author}
  {\bibfnamefont {A.}~\bibnamefont {Mazumdar}},\ }\bibfield  {title} {\bibinfo
  {title} {{Improving resilience of quantum-gravity-induced entanglement of
  masses to decoherence using three superpositions}},\ }\href
  {https://doi.org/10.1103/PhysRevA.105.032411} {\bibfield  {journal} {\bibinfo
   {journal} {Phys. Rev. A}\ }\textbf {\bibinfo {volume} {105}},\ \bibinfo
  {pages} {032411} (\bibinfo {year} {2022})},\ \Eprint
  {https://arxiv.org/abs/2110.14695} {arXiv:2110.14695 [quant-ph]} \BibitemShut
  {NoStop}%
\bibitem [{\citenamefont {Tilly}\ \emph {et~al.}(2021)\citenamefont {Tilly},
  \citenamefont {Marshman}, \citenamefont {Mazumdar},\ and\ \citenamefont
  {Bose}}]{Tilly:2021qef}%
  \BibitemOpen
  \bibfield  {author} {\bibinfo {author} {\bibfnamefont {J.}~\bibnamefont
  {Tilly}}, \bibinfo {author} {\bibfnamefont {R.~J.}\ \bibnamefont {Marshman}},
  \bibinfo {author} {\bibfnamefont {A.}~\bibnamefont {Mazumdar}},\ and\
  \bibinfo {author} {\bibfnamefont {S.}~\bibnamefont {Bose}},\ }\bibfield
  {title} {\bibinfo {title} {{Qudits for witnessing quantum-gravity-induced
  entanglement of masses under decoherence}},\ }\href
  {https://doi.org/10.1103/PhysRevA.104.052416} {\bibfield  {journal} {\bibinfo
   {journal} {Phys. Rev. A}\ }\textbf {\bibinfo {volume} {104}},\ \bibinfo
  {pages} {052416} (\bibinfo {year} {2021})},\ \Eprint
  {https://arxiv.org/abs/2101.08086} {arXiv:2101.08086 [quant-ph]} \BibitemShut
  {NoStop}%
\bibitem [{\citenamefont {Marletto}\ and\ \citenamefont
  {Vedral}(2017)}]{Marletto:2017kzi}%
  \BibitemOpen
  \bibfield  {author} {\bibinfo {author} {\bibfnamefont {C.}~\bibnamefont
  {Marletto}}\ and\ \bibinfo {author} {\bibfnamefont {V.}~\bibnamefont
  {Vedral}},\ }\bibfield  {title} {\bibinfo {title} {{Gravitationally-induced
  entanglement between two massive particles is sufficient evidence of quantum
  effects in gravity}},\ }\href
  {https://doi.org/10.1103/PhysRevLett.119.240402} {\bibfield  {journal}
  {\bibinfo  {journal} {Phys. Rev. Lett.}\ }\textbf {\bibinfo {volume} {119}},\
  \bibinfo {pages} {240402} (\bibinfo {year} {2017})},\ \Eprint
  {https://arxiv.org/abs/1707.06036} {arXiv:1707.06036 [quant-ph]} \BibitemShut
  {NoStop}%
\bibitem [{\citenamefont {van~de Kamp}\ \emph {et~al.}(2020)\citenamefont
  {van~de Kamp}, \citenamefont {Marshman}, \citenamefont {Bose},\ and\
  \citenamefont {Mazumdar}}]{vandeKamp:2020rqh}%
  \BibitemOpen
  \bibfield  {author} {\bibinfo {author} {\bibfnamefont {T.~W.}\ \bibnamefont
  {van~de Kamp}}, \bibinfo {author} {\bibfnamefont {R.~J.}\ \bibnamefont
  {Marshman}}, \bibinfo {author} {\bibfnamefont {S.}~\bibnamefont {Bose}},\
  and\ \bibinfo {author} {\bibfnamefont {A.}~\bibnamefont {Mazumdar}},\
  }\bibfield  {title} {\bibinfo {title} {{Quantum Gravity Witness via
  Entanglement of Masses: Casimir Screening}},\ }\href
  {https://doi.org/10.1103/PhysRevA.102.062807} {\bibfield  {journal} {\bibinfo
   {journal} {Phys. Rev. A}\ }\textbf {\bibinfo {volume} {102}},\ \bibinfo
  {pages} {062807} (\bibinfo {year} {2020})},\ \Eprint
  {https://arxiv.org/abs/2006.06931} {arXiv:2006.06931 [quant-ph]} \BibitemShut
  {NoStop}%
\bibitem [{\citenamefont {Marshman}\ \emph {et~al.}(2024)\citenamefont
  {Marshman}, \citenamefont {Bose}, \citenamefont {Geraci},\ and\ \citenamefont
  {Mazumdar}}]{marshman2024entanglement}%
  \BibitemOpen
  \bibfield  {author} {\bibinfo {author} {\bibfnamefont {R.~J.}\ \bibnamefont
  {Marshman}}, \bibinfo {author} {\bibfnamefont {S.}~\bibnamefont {Bose}},
  \bibinfo {author} {\bibfnamefont {A.}~\bibnamefont {Geraci}},\ and\ \bibinfo
  {author} {\bibfnamefont {A.}~\bibnamefont {Mazumdar}},\ }\bibfield  {title}
  {\bibinfo {title} {Entanglement of magnetically levitated massive
  schr{\"o}dinger cat states by induced dipole interaction},\ }\href
  {https://doi.org/10.1103/PhysRevA.109.L030401} {\bibfield  {journal}
  {\bibinfo  {journal} {Physical Review A}\ }\textbf {\bibinfo {volume}
  {109}},\ \bibinfo {pages} {L030401} (\bibinfo {year} {2024})}\BibitemShut
  {NoStop}%
\bibitem [{\citenamefont {Elahi}\ and\ \citenamefont
  {Mazumdar}(2023)}]{elahi2023probing}%
  \BibitemOpen
  \bibfield  {author} {\bibinfo {author} {\bibfnamefont {S.~G.}\ \bibnamefont
  {Elahi}}\ and\ \bibinfo {author} {\bibfnamefont {A.}~\bibnamefont
  {Mazumdar}},\ }\bibfield  {title} {\bibinfo {title} {Probing massless and
  massive gravitons via entanglement in a warped extra dimension},\ }\href
  {https://doi.org/10.1103/PhysRevD.108.035018} {\bibfield  {journal} {\bibinfo
   {journal} {Physical Review D}\ }\textbf {\bibinfo {volume} {108}},\ \bibinfo
  {pages} {035018} (\bibinfo {year} {2023})}\BibitemShut {NoStop}%
\bibitem [{\citenamefont {Chakraborty}\ \emph {et~al.}(2023)\citenamefont
  {Chakraborty}, \citenamefont {Mazumdar},\ and\ \citenamefont
  {Pradhan}}]{chakraborty2023distinguishing}%
  \BibitemOpen
  \bibfield  {author} {\bibinfo {author} {\bibfnamefont {S.}~\bibnamefont
  {Chakraborty}}, \bibinfo {author} {\bibfnamefont {A.}~\bibnamefont
  {Mazumdar}},\ and\ \bibinfo {author} {\bibfnamefont {R.}~\bibnamefont
  {Pradhan}},\ }\bibfield  {title} {\bibinfo {title} {Distinguishing jordan and
  einstein frames in gravity through entanglement},\ }\href
  {https://doi.org/10.1103/PhysRevD.108.L121505} {\bibfield  {journal}
  {\bibinfo  {journal} {Physical Review D}\ }\textbf {\bibinfo {volume}
  {108}},\ \bibinfo {pages} {L121505} (\bibinfo {year} {2023})}\BibitemShut
  {NoStop}%
\bibitem [{\citenamefont {Toro\v{s}}\ \emph
  {et~al.}(2024{\natexlab{a}})\citenamefont {Toro\v{s}}, \citenamefont {Schut},
  \citenamefont {Andriolo}, \citenamefont {Bose},\ and\ \citenamefont
  {Mazumdar}}]{Toros:2024ozf}%
  \BibitemOpen
  \bibfield  {author} {\bibinfo {author} {\bibfnamefont {M.}~\bibnamefont
  {Toro\v{s}}}, \bibinfo {author} {\bibfnamefont {M.}~\bibnamefont {Schut}},
  \bibinfo {author} {\bibfnamefont {P.}~\bibnamefont {Andriolo}}, \bibinfo
  {author} {\bibfnamefont {S.}~\bibnamefont {Bose}},\ and\ \bibinfo {author}
  {\bibfnamefont {A.}~\bibnamefont {Mazumdar}},\ }\bibfield  {title} {\bibinfo
  {title} {{Relativistic Dips in Entangling Power of Gravity}},\ }\href@noop {}
  {\  (\bibinfo {year} {2024}{\natexlab{a}})},\ \Eprint
  {https://arxiv.org/abs/2405.04661} {arXiv:2405.04661 [quant-ph]} \BibitemShut
  {NoStop}%
\bibitem [{\citenamefont {Toro\v{s}}\ \emph
  {et~al.}(2024{\natexlab{b}})\citenamefont {Toro\v{s}}, \citenamefont
  {Andriolo}, \citenamefont {Schut}, \citenamefont {Bose},\ and\ \citenamefont
  {Mazumdar}}]{Toros:2024ozu}%
  \BibitemOpen
  \bibfield  {author} {\bibinfo {author} {\bibfnamefont {M.}~\bibnamefont
  {Toro\v{s}}}, \bibinfo {author} {\bibfnamefont {P.}~\bibnamefont {Andriolo}},
  \bibinfo {author} {\bibfnamefont {M.}~\bibnamefont {Schut}}, \bibinfo
  {author} {\bibfnamefont {S.}~\bibnamefont {Bose}},\ and\ \bibinfo {author}
  {\bibfnamefont {A.}~\bibnamefont {Mazumdar}},\ }\bibfield  {title} {\bibinfo
  {title} {{Relativistic Effects on Entangled Single-Electron Traps}},\
  }\href@noop {} {\  (\bibinfo {year} {2024}{\natexlab{b}})},\ \Eprint
  {https://arxiv.org/abs/2406.17848} {arXiv:2406.17848 [quant-ph]} \BibitemShut
  {NoStop}%
\bibitem [{\citenamefont {Beckering~Vinckers}\ \emph
  {et~al.}(2023)\citenamefont {Beckering~Vinckers}, \citenamefont {De~La
  Cruz-Dombriz},\ and\ \citenamefont {Mazumdar}}]{Vinckers:2023grv}%
  \BibitemOpen
  \bibfield  {author} {\bibinfo {author} {\bibfnamefont {U.~K.}\ \bibnamefont
  {Beckering~Vinckers}}, \bibinfo {author} {\bibfnamefont {{\'A}.}~\bibnamefont
  {De~La Cruz-Dombriz}},\ and\ \bibinfo {author} {\bibfnamefont
  {A.}~\bibnamefont {Mazumdar}},\ }\bibfield  {title} {\bibinfo {title}
  {Quantum entanglement of masses with nonlocal gravitational interaction},\
  }\href {https://doi.org/10.1103/PhysRevD.107.124036} {\bibfield  {journal}
  {\bibinfo  {journal} {Physical Review D}\ }\textbf {\bibinfo {volume}
  {107}},\ \bibinfo {pages} {124036} (\bibinfo {year} {2023})}\BibitemShut
  {NoStop}%
\bibitem [{\citenamefont {Schut}\ \emph {et~al.}(2024)\citenamefont {Schut},
  \citenamefont {Geraci}, \citenamefont {Bose},\ and\ \citenamefont
  {Mazumdar}}]{Schut:2023hsy}%
  \BibitemOpen
  \bibfield  {author} {\bibinfo {author} {\bibfnamefont {M.}~\bibnamefont
  {Schut}}, \bibinfo {author} {\bibfnamefont {A.}~\bibnamefont {Geraci}},
  \bibinfo {author} {\bibfnamefont {S.}~\bibnamefont {Bose}},\ and\ \bibinfo
  {author} {\bibfnamefont {A.}~\bibnamefont {Mazumdar}},\ }\bibfield  {title}
  {\bibinfo {title} {{Micrometer-size spatial superpositions for the QGEM
  protocol via screening and trapping}},\ }\href
  {https://doi.org/10.1103/PhysRevResearch.6.013199} {\bibfield  {journal}
  {\bibinfo  {journal} {Phys. Rev. Res.}\ }\textbf {\bibinfo {volume} {6}},\
  \bibinfo {pages} {013199} (\bibinfo {year} {2024})},\ \Eprint
  {https://arxiv.org/abs/2307.15743} {arXiv:2307.15743 [quant-ph]} \BibitemShut
  {NoStop}%
\bibitem [{\citenamefont {Schut}\ \emph
  {et~al.}(2023{\natexlab{a}})\citenamefont {Schut}, \citenamefont {Grinin},
  \citenamefont {Dana}, \citenamefont {Bose}, \citenamefont {Geraci},\ and\
  \citenamefont {Mazumdar}}]{Schut:2023eux}%
  \BibitemOpen
  \bibfield  {author} {\bibinfo {author} {\bibfnamefont {M.}~\bibnamefont
  {Schut}}, \bibinfo {author} {\bibfnamefont {A.}~\bibnamefont {Grinin}},
  \bibinfo {author} {\bibfnamefont {A.}~\bibnamefont {Dana}}, \bibinfo {author}
  {\bibfnamefont {S.}~\bibnamefont {Bose}}, \bibinfo {author} {\bibfnamefont
  {A.}~\bibnamefont {Geraci}},\ and\ \bibinfo {author} {\bibfnamefont
  {A.}~\bibnamefont {Mazumdar}},\ }\bibfield  {title} {\bibinfo {title}
  {{Relaxation of experimental parameters in a quantum-gravity-induced
  entanglement of masses protocol using electromagnetic screening}},\ }\href
  {https://doi.org/10.1103/PhysRevResearch.5.043170} {\bibfield  {journal}
  {\bibinfo  {journal} {Phys. Rev. Res.}\ }\textbf {\bibinfo {volume} {5}},\
  \bibinfo {pages} {043170} (\bibinfo {year} {2023}{\natexlab{a}})},\ \Eprint
  {https://arxiv.org/abs/2307.07536} {arXiv:2307.07536 [quant-ph]} \BibitemShut
  {NoStop}%
\bibitem [{\citenamefont {Fein}\ \emph {et~al.}(2019)\citenamefont {Fein} \emph
  {et~al.}}]{arndt}%
  \BibitemOpen
  \bibfield  {author} {\bibinfo {author} {\bibfnamefont {Y.~Y.}\ \bibnamefont
  {Fein}} \emph {et~al.},\ }\bibfield  {title} {\bibinfo {title} {Quantum
  superposition of molecules beyond 25 kda},\ }\href
  {https://doi.org/10.1038/s41567-019-0663-9} {\bibfield  {journal} {\bibinfo
  {journal} {Nature Phys.}\ }\textbf {\bibinfo {volume} {15}},\ \bibinfo
  {pages} {1242} (\bibinfo {year} {2019})}\BibitemShut {NoStop}%
\bibitem [{\citenamefont {Overstreet}\ \emph {et~al.}(2022)\citenamefont
  {Overstreet}, \citenamefont {Asenbaum}, \citenamefont {Curti}, \citenamefont
  {Kim},\ and\ \citenamefont {Kasevich}}]{overstreet2022observation}%
  \BibitemOpen
  \bibfield  {author} {\bibinfo {author} {\bibfnamefont {C.}~\bibnamefont
  {Overstreet}}, \bibinfo {author} {\bibfnamefont {P.}~\bibnamefont
  {Asenbaum}}, \bibinfo {author} {\bibfnamefont {J.}~\bibnamefont {Curti}},
  \bibinfo {author} {\bibfnamefont {M.}~\bibnamefont {Kim}},\ and\ \bibinfo
  {author} {\bibfnamefont {M.~A.}\ \bibnamefont {Kasevich}},\ }\bibfield
  {title} {\bibinfo {title} {Observation of a gravitational aharonov-bohm
  effect},\ }\href {https://doi.org/10.1126/science.abl7152} {\bibfield
  {journal} {\bibinfo  {journal} {Science}\ }\textbf {\bibinfo {volume}
  {375}},\ \bibinfo {pages} {226} (\bibinfo {year} {2022})}\BibitemShut
  {NoStop}%
\bibitem [{\citenamefont {Asenbaum}\ \emph {et~al.}(2017)\citenamefont
  {Asenbaum}, \citenamefont {Overstreet}, \citenamefont {Kovachy},
  \citenamefont {Brown}, \citenamefont {Hogan},\ and\ \citenamefont
  {Kasevich}}]{asenbaum2017phase}%
  \BibitemOpen
  \bibfield  {author} {\bibinfo {author} {\bibfnamefont {P.}~\bibnamefont
  {Asenbaum}}, \bibinfo {author} {\bibfnamefont {C.}~\bibnamefont
  {Overstreet}}, \bibinfo {author} {\bibfnamefont {T.}~\bibnamefont {Kovachy}},
  \bibinfo {author} {\bibfnamefont {D.~D.}\ \bibnamefont {Brown}}, \bibinfo
  {author} {\bibfnamefont {J.~M.}\ \bibnamefont {Hogan}},\ and\ \bibinfo
  {author} {\bibfnamefont {M.~A.}\ \bibnamefont {Kasevich}},\ }\bibfield
  {title} {\bibinfo {title} {Phase shift in an atom interferometer due to
  spacetime curvature across its wave function},\ }\href
  {https://doi.org/10.1103/PhysRevLett.118.183602} {\bibfield  {journal}
  {\bibinfo  {journal} {Physical review letters}\ }\textbf {\bibinfo {volume}
  {118}},\ \bibinfo {pages} {183602} (\bibinfo {year} {2017})}\BibitemShut
  {NoStop}%
\bibitem [{\citenamefont {Rijavec}\ \emph {et~al.}(2021)\citenamefont
  {Rijavec}, \citenamefont {Carlesso}, \citenamefont {Bassi}, \citenamefont
  {Vedral},\ and\ \citenamefont {Marletto}}]{Rijavec:2020qxd}%
  \BibitemOpen
  \bibfield  {author} {\bibinfo {author} {\bibfnamefont {S.}~\bibnamefont
  {Rijavec}}, \bibinfo {author} {\bibfnamefont {M.}~\bibnamefont {Carlesso}},
  \bibinfo {author} {\bibfnamefont {A.}~\bibnamefont {Bassi}}, \bibinfo
  {author} {\bibfnamefont {V.}~\bibnamefont {Vedral}},\ and\ \bibinfo {author}
  {\bibfnamefont {C.}~\bibnamefont {Marletto}},\ }\bibfield  {title} {\bibinfo
  {title} {{Decoherence effects in non-classicality tests of gravity}},\ }\href
  {https://doi.org/10.1088/1367-2630/abf3eb} {\bibfield  {journal} {\bibinfo
  {journal} {New J. Phys.}\ }\textbf {\bibinfo {volume} {23}},\ \bibinfo
  {pages} {043040} (\bibinfo {year} {2021})},\ \Eprint
  {https://arxiv.org/abs/2012.06230} {arXiv:2012.06230 [quant-ph]} \BibitemShut
  {NoStop}%
\bibitem [{\citenamefont {Fragolino}\ \emph {et~al.}(2024)\citenamefont
  {Fragolino}, \citenamefont {Schut}, \citenamefont {Toro\v{s}}, \citenamefont
  {Bose},\ and\ \citenamefont {Mazumdar}}]{Fragolino:2023agd}%
  \BibitemOpen
  \bibfield  {author} {\bibinfo {author} {\bibfnamefont {P.}~\bibnamefont
  {Fragolino}}, \bibinfo {author} {\bibfnamefont {M.}~\bibnamefont {Schut}},
  \bibinfo {author} {\bibfnamefont {M.}~\bibnamefont {Toro\v{s}}}, \bibinfo
  {author} {\bibfnamefont {S.}~\bibnamefont {Bose}},\ and\ \bibinfo {author}
  {\bibfnamefont {A.}~\bibnamefont {Mazumdar}},\ }\bibfield  {title} {\bibinfo
  {title} {{Decoherence of a matter-wave interferometer due to dipole-dipole
  interactions}},\ }\href {https://doi.org/10.1103/PhysRevA.109.033301}
  {\bibfield  {journal} {\bibinfo  {journal} {Phys. Rev. A}\ }\textbf {\bibinfo
  {volume} {109}},\ \bibinfo {pages} {033301} (\bibinfo {year} {2024})},\
  \Eprint {https://arxiv.org/abs/2307.07001} {arXiv:2307.07001 [quant-ph]}
  \BibitemShut {NoStop}%
\bibitem [{\citenamefont {Schut}\ \emph
  {et~al.}(2023{\natexlab{b}})\citenamefont {Schut}, \citenamefont {Bosma},
  \citenamefont {Wu}, \citenamefont {Toro\v{s}}, \citenamefont {Bose},\ and\
  \citenamefont {Mazumdar}}]{Schut:2023tce}%
  \BibitemOpen
  \bibfield  {author} {\bibinfo {author} {\bibfnamefont {M.}~\bibnamefont
  {Schut}}, \bibinfo {author} {\bibfnamefont {H.}~\bibnamefont {Bosma}},
  \bibinfo {author} {\bibfnamefont {M.}~\bibnamefont {Wu}}, \bibinfo {author}
  {\bibfnamefont {M.}~\bibnamefont {Toro\v{s}}}, \bibinfo {author}
  {\bibfnamefont {S.}~\bibnamefont {Bose}},\ and\ \bibinfo {author}
  {\bibfnamefont {A.}~\bibnamefont {Mazumdar}},\ }\bibfield  {title} {\bibinfo
  {title} {{Dephasing due to electromagnetic interactions in spatial qubits}},\
  }\href@noop {} {\  (\bibinfo {year} {2023}{\natexlab{b}})},\ \Eprint
  {https://arxiv.org/abs/2312.05452} {arXiv:2312.05452 [quant-ph]} \BibitemShut
  {NoStop}%
\bibitem [{\citenamefont {Zhou}\ \emph
  {et~al.}(2022{\natexlab{a}})\citenamefont {Zhou}, \citenamefont {Marshman},
  \citenamefont {Bose},\ and\ \citenamefont {Mazumdar}}]{Zhou:2022epb}%
  \BibitemOpen
  \bibfield  {author} {\bibinfo {author} {\bibfnamefont {R.}~\bibnamefont
  {Zhou}}, \bibinfo {author} {\bibfnamefont {R.~J.}\ \bibnamefont {Marshman}},
  \bibinfo {author} {\bibfnamefont {S.}~\bibnamefont {Bose}},\ and\ \bibinfo
  {author} {\bibfnamefont {A.}~\bibnamefont {Mazumdar}},\ }\bibfield  {title}
  {\bibinfo {title} {{Gravito-diamagnetic forces for mass independent large
  spatial quantum superpositions}},\ }\href@noop {} {\  (\bibinfo {year}
  {2022}{\natexlab{a}})},\ \Eprint {https://arxiv.org/abs/2211.08435}
  {arXiv:2211.08435 [quant-ph]} \BibitemShut {NoStop}%
\bibitem [{\citenamefont {Zhou}\ \emph
  {et~al.}(2022{\natexlab{b}})\citenamefont {Zhou}, \citenamefont {Marshman},
  \citenamefont {Bose},\ and\ \citenamefont {Mazumdar}}]{Zhou:2022frl}%
  \BibitemOpen
  \bibfield  {author} {\bibinfo {author} {\bibfnamefont {R.}~\bibnamefont
  {Zhou}}, \bibinfo {author} {\bibfnamefont {R.~J.}\ \bibnamefont {Marshman}},
  \bibinfo {author} {\bibfnamefont {S.}~\bibnamefont {Bose}},\ and\ \bibinfo
  {author} {\bibfnamefont {A.}~\bibnamefont {Mazumdar}},\ }\bibfield  {title}
  {\bibinfo {title} {{Catapulting towards massive and large spatial quantum
  superposition}},\ }\href {https://doi.org/10.1103/PhysRevResearch.4.043157}
  {\bibfield  {journal} {\bibinfo  {journal} {Phys. Rev. Res.}\ }\textbf
  {\bibinfo {volume} {4}},\ \bibinfo {pages} {043157} (\bibinfo {year}
  {2022}{\natexlab{b}})},\ \Eprint {https://arxiv.org/abs/2206.04088}
  {arXiv:2206.04088 [quant-ph]} \BibitemShut {NoStop}%
\bibitem [{\citenamefont {Zhou}\ \emph {et~al.}(2023)\citenamefont {Zhou},
  \citenamefont {Marshman}, \citenamefont {Bose},\ and\ \citenamefont
  {Mazumdar}}]{Zhou:2022jug}%
  \BibitemOpen
  \bibfield  {author} {\bibinfo {author} {\bibfnamefont {R.}~\bibnamefont
  {Zhou}}, \bibinfo {author} {\bibfnamefont {R.~J.}\ \bibnamefont {Marshman}},
  \bibinfo {author} {\bibfnamefont {S.}~\bibnamefont {Bose}},\ and\ \bibinfo
  {author} {\bibfnamefont {A.}~\bibnamefont {Mazumdar}},\ }\bibfield  {title}
  {\bibinfo {title} {{Mass-independent scheme for enhancing spatial quantum
  superpositions}},\ }\href {https://doi.org/10.1103/PhysRevA.107.032212}
  {\bibfield  {journal} {\bibinfo  {journal} {Phys. Rev. A}\ }\textbf {\bibinfo
  {volume} {107}},\ \bibinfo {pages} {032212} (\bibinfo {year} {2023})},\
  \Eprint {https://arxiv.org/abs/2210.05689} {arXiv:2210.05689 [quant-ph]}
  \BibitemShut {NoStop}%
\bibitem [{\citenamefont {Englert}\ \emph {et~al.}(1988)\citenamefont
  {Englert}, \citenamefont {Schwinger},\ and\ \citenamefont
  {Scully}}]{Englert}%
  \BibitemOpen
  \bibfield  {author} {\bibinfo {author} {\bibfnamefont {B.}~\bibnamefont
  {Englert}}, \bibinfo {author} {\bibfnamefont {J.}~\bibnamefont {Schwinger}},\
  and\ \bibinfo {author} {\bibfnamefont {M.~O.}\ \bibnamefont {Scully}},\
  }\bibfield  {title} {\bibinfo {title} {Is spin coherence like humpty-dumpty?
  i. simplified treatment},\ }\href {https://doi.org/10.1007/bf01909939}
  {\bibfield  {journal} {\bibinfo  {journal} {Foundations of Physics}\ }\textbf
  {\bibinfo {volume} {18}},\ \bibinfo {pages} {1045} (\bibinfo {year}
  {1988})}\BibitemShut {NoStop}%
\bibitem [{\citenamefont {{Schwinger}}\ \emph {et~al.}(1988)\citenamefont
  {{Schwinger}}, \citenamefont {{Scully}},\ and\ \citenamefont
  {{Englert}}}]{Schwinger}%
  \BibitemOpen
  \bibfield  {author} {\bibinfo {author} {\bibfnamefont {J.}~\bibnamefont
  {{Schwinger}}}, \bibinfo {author} {\bibfnamefont {M.~O.}\ \bibnamefont
  {{Scully}}},\ and\ \bibinfo {author} {\bibfnamefont {B.~G.}\ \bibnamefont
  {{Englert}}},\ }\bibfield  {title} {\bibinfo {title} {{Is spin coherence like
  Humpty-Dumpty?}},\ }\href {https://doi.org/10.1007/BF01384847} {\bibfield
  {journal} {\bibinfo  {journal} {Zeitschrift fur Physik D Atoms Molecules
  Clusters}\ }\textbf {\bibinfo {volume} {10}},\ \bibinfo {pages} {135}
  (\bibinfo {year} {1988})}\BibitemShut {NoStop}%
\bibitem [{\citenamefont {Japha}\ and\ \citenamefont
  {Folman}(2023)}]{Japha:2022phw}%
  \BibitemOpen
  \bibfield  {author} {\bibinfo {author} {\bibfnamefont {Y.}~\bibnamefont
  {Japha}}\ and\ \bibinfo {author} {\bibfnamefont {R.}~\bibnamefont {Folman}},\
  }\bibfield  {title} {\bibinfo {title} {{Quantum Uncertainty Limit for
  Stern-Gerlach Interferometry with Massive Objects}},\ }\href
  {https://doi.org/10.1103/PhysRevLett.130.113602} {\bibfield  {journal}
  {\bibinfo  {journal} {Phys. Rev. Lett.}\ }\textbf {\bibinfo {volume} {130}},\
  \bibinfo {pages} {113602} (\bibinfo {year} {2023})},\ \Eprint
  {https://arxiv.org/abs/2202.10535} {arXiv:2202.10535 [quant-ph]} \BibitemShut
  {NoStop}%
\bibitem [{\citenamefont {Japha}\ and\ \citenamefont
  {Folman}(2022)}]{japha2022role}%
  \BibitemOpen
  \bibfield  {author} {\bibinfo {author} {\bibfnamefont {Y.}~\bibnamefont
  {Japha}}\ and\ \bibinfo {author} {\bibfnamefont {R.}~\bibnamefont {Folman}},\
  }\bibfield  {title} {\bibinfo {title} {Role of rotations in stern-gerlach
  interferometry with massive objects},\ }\href@noop {} {\  (\bibinfo {year}
  {2022})},\ \Eprint {https://arxiv.org/abs/2202.10535} {arXiv:2202.10535
  [quant-ph]} \BibitemShut {NoStop}%
\bibitem [{\citenamefont {Afek}\ \emph {et~al.}(2021)\citenamefont {Afek},
  \citenamefont {Monteiro}, \citenamefont {Siegel}, \citenamefont {Wang},
  \citenamefont {Dickson}, \citenamefont {Recoaro}, \citenamefont {Watts},\
  and\ \citenamefont {Moore}}]{Afek:2021bua}%
  \BibitemOpen
  \bibfield  {author} {\bibinfo {author} {\bibfnamefont {G.}~\bibnamefont
  {Afek}}, \bibinfo {author} {\bibfnamefont {F.}~\bibnamefont {Monteiro}},
  \bibinfo {author} {\bibfnamefont {B.}~\bibnamefont {Siegel}}, \bibinfo
  {author} {\bibfnamefont {J.}~\bibnamefont {Wang}}, \bibinfo {author}
  {\bibfnamefont {S.}~\bibnamefont {Dickson}}, \bibinfo {author} {\bibfnamefont
  {J.}~\bibnamefont {Recoaro}}, \bibinfo {author} {\bibfnamefont
  {M.}~\bibnamefont {Watts}},\ and\ \bibinfo {author} {\bibfnamefont {D.~C.}\
  \bibnamefont {Moore}},\ }\bibfield  {title} {\bibinfo {title} {{Control and
  measurement of electric dipole moments in levitated optomechanics}},\ }\href
  {https://doi.org/10.1103/PhysRevA.104.053512} {\bibfield  {journal} {\bibinfo
   {journal} {Phys. Rev. A}\ }\textbf {\bibinfo {volume} {104}},\ \bibinfo
  {pages} {053512} (\bibinfo {year} {2021})},\ \Eprint
  {https://arxiv.org/abs/2108.04406} {arXiv:2108.04406 [physics.optics]}
  \BibitemShut {NoStop}%
\bibitem [{\citenamefont {Deli{\'{c}}}\ \emph {et~al.}(2020)\citenamefont
  {Deli{\'{c}}}, \citenamefont {Reisenbauer}, \citenamefont {Dare},
  \citenamefont {Grass}, \citenamefont {Vuleti{\'{c}}}, \citenamefont
  {Kiesel},\ and\ \citenamefont {Aspelmeyer}}]{Deli__2020}%
  \BibitemOpen
  \bibfield  {author} {\bibinfo {author} {\bibfnamefont {U.}~\bibnamefont
  {Deli{\'{c}}}}, \bibinfo {author} {\bibfnamefont {M.}~\bibnamefont
  {Reisenbauer}}, \bibinfo {author} {\bibfnamefont {K.}~\bibnamefont {Dare}},
  \bibinfo {author} {\bibfnamefont {D.}~\bibnamefont {Grass}}, \bibinfo
  {author} {\bibfnamefont {V.}~\bibnamefont {Vuleti{\'{c}}}}, \bibinfo {author}
  {\bibfnamefont {N.}~\bibnamefont {Kiesel}},\ and\ \bibinfo {author}
  {\bibfnamefont {M.}~\bibnamefont {Aspelmeyer}},\ }\bibfield  {title}
  {\bibinfo {title} {Cooling of a levitated nanoparticle to the motional
  quantum ground state},\ }\href {https://doi.org/10.1126/science.aba3993}
  {\bibfield  {journal} {\bibinfo  {journal} {Science}\ }\textbf {\bibinfo
  {volume} {367}},\ \bibinfo {pages} {892} (\bibinfo {year}
  {2020})}\BibitemShut {NoStop}%
\bibitem [{\citenamefont {Gieseler}\ \emph {et~al.}(2012)\citenamefont
  {Gieseler}, \citenamefont {Deutsch}, \citenamefont {Quidant},\ and\
  \citenamefont {Novotny}}]{gieseler2012subkelvin}%
  \BibitemOpen
  \bibfield  {author} {\bibinfo {author} {\bibfnamefont {J.}~\bibnamefont
  {Gieseler}}, \bibinfo {author} {\bibfnamefont {B.}~\bibnamefont {Deutsch}},
  \bibinfo {author} {\bibfnamefont {R.}~\bibnamefont {Quidant}},\ and\ \bibinfo
  {author} {\bibfnamefont {L.}~\bibnamefont {Novotny}},\ }\bibfield  {title}
  {\bibinfo {title} {Subkelvin parametric feedback cooling of a laser-trapped
  nanoparticle},\ }\href
  {https://doi.org/https://doi.org/10.1103/PhysRevLett.109.103603} {\bibfield
  {journal} {\bibinfo  {journal} {Physical review letters}\ }\textbf {\bibinfo
  {volume} {109}},\ \bibinfo {pages} {103603} (\bibinfo {year}
  {2012})}\BibitemShut {NoStop}%
\bibitem [{\citenamefont {Piotrowski}\ \emph {et~al.}(2023)\citenamefont
  {Piotrowski}, \citenamefont {Windey}, \citenamefont {Vijayan}, \citenamefont
  {Gonzalez-Ballestero}, \citenamefont {de~los R{\'{\i}}os~Sommer},
  \citenamefont {Meyer}, \citenamefont {Quidant}, \citenamefont {Romero-Isart},
  \citenamefont {Reimann},\ and\ \citenamefont {Novotny}}]{Piotrowski_2023}%
  \BibitemOpen
  \bibfield  {author} {\bibinfo {author} {\bibfnamefont {J.}~\bibnamefont
  {Piotrowski}}, \bibinfo {author} {\bibfnamefont {D.}~\bibnamefont {Windey}},
  \bibinfo {author} {\bibfnamefont {J.}~\bibnamefont {Vijayan}}, \bibinfo
  {author} {\bibfnamefont {C.}~\bibnamefont {Gonzalez-Ballestero}}, \bibinfo
  {author} {\bibfnamefont {A.}~\bibnamefont {de~los R{\'{\i}}os~Sommer}},
  \bibinfo {author} {\bibfnamefont {N.}~\bibnamefont {Meyer}}, \bibinfo
  {author} {\bibfnamefont {R.}~\bibnamefont {Quidant}}, \bibinfo {author}
  {\bibfnamefont {O.}~\bibnamefont {Romero-Isart}}, \bibinfo {author}
  {\bibfnamefont {R.}~\bibnamefont {Reimann}},\ and\ \bibinfo {author}
  {\bibfnamefont {L.}~\bibnamefont {Novotny}},\ }\bibfield  {title} {\bibinfo
  {title} {Simultaneous ground-state cooling of two mechanical modes of a
  levitated nanoparticle},\ }\href {https://doi.org/10.1038/s41567-023-01956-1}
  {\bibfield  {journal} {\bibinfo  {journal} {Nature Physics}\ }\textbf
  {\bibinfo {volume} {19}},\ \bibinfo {pages} {1009} (\bibinfo {year}
  {2023})}\BibitemShut {NoStop}%
\bibitem [{\citenamefont {Abobeih}\ \emph {et~al.}(2018)\citenamefont
  {Abobeih}, \citenamefont {Cramer}, \citenamefont {Bakker}, \citenamefont
  {Kalb}, \citenamefont {Markham}, \citenamefont {Twitchen},\ and\
  \citenamefont {Taminiau}}]{Abobeih_2018}%
  \BibitemOpen
  \bibfield  {author} {\bibinfo {author} {\bibfnamefont {M.~H.}\ \bibnamefont
  {Abobeih}}, \bibinfo {author} {\bibfnamefont {J.}~\bibnamefont {Cramer}},
  \bibinfo {author} {\bibfnamefont {M.~A.}\ \bibnamefont {Bakker}}, \bibinfo
  {author} {\bibfnamefont {N.}~\bibnamefont {Kalb}}, \bibinfo {author}
  {\bibfnamefont {M.}~\bibnamefont {Markham}}, \bibinfo {author} {\bibfnamefont
  {D.~J.}\ \bibnamefont {Twitchen}},\ and\ \bibinfo {author} {\bibfnamefont
  {T.~H.}\ \bibnamefont {Taminiau}},\ }\bibfield  {title} {\bibinfo {title}
  {One-second coherence for a single electron spin coupled to a multi-qubit
  nuclear-spin environment},\ }\bibfield  {journal} {\bibinfo  {journal}
  {Nature Communications}\ }\textbf {\bibinfo {volume} {9}},\ \href
  {https://doi.org/10.1038/s41467-018-04916-z} {10.1038/s41467-018-04916-z}
  (\bibinfo {year} {2018})\BibitemShut {NoStop}%
\bibitem [{\citenamefont {Friese}\ \emph {et~al.}(1998)\citenamefont {Friese},
  \citenamefont {Nieminen}, \citenamefont {Heckenberg},\ and\ \citenamefont
  {Rubinsztein-Dunlop}}]{friese1998optical}%
  \BibitemOpen
  \bibfield  {author} {\bibinfo {author} {\bibfnamefont {M.~E.}\ \bibnamefont
  {Friese}}, \bibinfo {author} {\bibfnamefont {T.~A.}\ \bibnamefont
  {Nieminen}}, \bibinfo {author} {\bibfnamefont {N.~R.}\ \bibnamefont
  {Heckenberg}},\ and\ \bibinfo {author} {\bibfnamefont {H.}~\bibnamefont
  {Rubinsztein-Dunlop}},\ }\bibfield  {title} {\bibinfo {title} {Optical
  alignment and spinning of laser-trapped microscopic particles},\ }\href
  {https://doi.org/10.1038/28566} {\bibfield  {journal} {\bibinfo  {journal}
  {Nature}\ }\textbf {\bibinfo {volume} {394}},\ \bibinfo {pages} {348}
  (\bibinfo {year} {1998})}\BibitemShut {NoStop}%
\bibitem [{\citenamefont {Jin}\ \emph {et~al.}(2024)\citenamefont {Jin},
  \citenamefont {Shen}, \citenamefont {Ju}, \citenamefont {Gao}, \citenamefont
  {Zu}, \citenamefont {Grine},\ and\ \citenamefont {Li}}]{jin2024quantum}%
  \BibitemOpen
  \bibfield  {author} {\bibinfo {author} {\bibfnamefont {Y.}~\bibnamefont
  {Jin}}, \bibinfo {author} {\bibfnamefont {K.}~\bibnamefont {Shen}}, \bibinfo
  {author} {\bibfnamefont {P.}~\bibnamefont {Ju}}, \bibinfo {author}
  {\bibfnamefont {X.}~\bibnamefont {Gao}}, \bibinfo {author} {\bibfnamefont
  {C.}~\bibnamefont {Zu}}, \bibinfo {author} {\bibfnamefont {A.~J.}\
  \bibnamefont {Grine}},\ and\ \bibinfo {author} {\bibfnamefont
  {T.}~\bibnamefont {Li}},\ }\bibfield  {title} {\bibinfo {title} {Quantum
  control and berry phase of electron spins in rotating levitated diamonds in
  high vacuum},\ }\href {https://doi.org/10.1038/s41467-024-49175-3} {\bibfield
   {journal} {\bibinfo  {journal} {Nature Communications}\ }\textbf {\bibinfo
  {volume} {15}},\ \bibinfo {pages} {5063} (\bibinfo {year}
  {2024})}\BibitemShut {NoStop}%
\bibitem [{\citenamefont {Perdriat}\ \emph {et~al.}(2023)\citenamefont
  {Perdriat}, \citenamefont {Rusconi}, \citenamefont {Delord}, \citenamefont
  {Huillery}, \citenamefont {Pellet-Mary}, \citenamefont {Stickler},\ and\
  \citenamefont {H{\'e}tet}}]{perdriat2023spin}%
  \BibitemOpen
  \bibfield  {author} {\bibinfo {author} {\bibfnamefont {M.}~\bibnamefont
  {Perdriat}}, \bibinfo {author} {\bibfnamefont {C.~C.}\ \bibnamefont
  {Rusconi}}, \bibinfo {author} {\bibfnamefont {T.}~\bibnamefont {Delord}},
  \bibinfo {author} {\bibfnamefont {P.}~\bibnamefont {Huillery}}, \bibinfo
  {author} {\bibfnamefont {C.}~\bibnamefont {Pellet-Mary}}, \bibinfo {author}
  {\bibfnamefont {B.~A.}\ \bibnamefont {Stickler}},\ and\ \bibinfo {author}
  {\bibfnamefont {G.}~\bibnamefont {H{\'e}tet}},\ }\bibfield  {title} {\bibinfo
  {title} {Spin read-out of the motion of levitated electrically rotated
  diamonds},\ }\href@noop {} {\  (\bibinfo {year} {2023})},\ \Eprint
  {https://arxiv.org/abs/2309.01545} {arXiv:2309.01545 [quant-ph]} \BibitemShut
  {NoStop}%
\bibitem [{\citenamefont {Kuhn}\ \emph {et~al.}(2017)\citenamefont {Kuhn},
  \citenamefont {Stickler}, \citenamefont {Kosloff}, \citenamefont {Patolsky},
  \citenamefont {Hornberger}, \citenamefont {Arndt},\ and\ \citenamefont
  {Millen}}]{kuhn2017optically}%
  \BibitemOpen
  \bibfield  {author} {\bibinfo {author} {\bibfnamefont {S.}~\bibnamefont
  {Kuhn}}, \bibinfo {author} {\bibfnamefont {B.~A.}\ \bibnamefont {Stickler}},
  \bibinfo {author} {\bibfnamefont {A.}~\bibnamefont {Kosloff}}, \bibinfo
  {author} {\bibfnamefont {F.}~\bibnamefont {Patolsky}}, \bibinfo {author}
  {\bibfnamefont {K.}~\bibnamefont {Hornberger}}, \bibinfo {author}
  {\bibfnamefont {M.}~\bibnamefont {Arndt}},\ and\ \bibinfo {author}
  {\bibfnamefont {J.}~\bibnamefont {Millen}},\ }\bibfield  {title} {\bibinfo
  {title} {Optically driven ultra-stable nanomechanical rotor},\ }\href
  {https://doi.org/10.1038/s41467-017-01902-9} {\bibfield  {journal} {\bibinfo
  {journal} {Nature communications}\ }\textbf {\bibinfo {volume} {8}},\
  \bibinfo {pages} {1670} (\bibinfo {year} {2017})}\BibitemShut {NoStop}%
\bibitem [{\citenamefont {Reimann}\ \emph {et~al.}(2018)\citenamefont
  {Reimann}, \citenamefont {Doderer}, \citenamefont {Hebestreit}, \citenamefont
  {Diehl}, \citenamefont {Frimmer}, \citenamefont {Windey}, \citenamefont
  {Tebbenjohanns},\ and\ \citenamefont {Novotny}}]{reimann2018ghz}%
  \BibitemOpen
  \bibfield  {author} {\bibinfo {author} {\bibfnamefont {R.}~\bibnamefont
  {Reimann}}, \bibinfo {author} {\bibfnamefont {M.}~\bibnamefont {Doderer}},
  \bibinfo {author} {\bibfnamefont {E.}~\bibnamefont {Hebestreit}}, \bibinfo
  {author} {\bibfnamefont {R.}~\bibnamefont {Diehl}}, \bibinfo {author}
  {\bibfnamefont {M.}~\bibnamefont {Frimmer}}, \bibinfo {author} {\bibfnamefont
  {D.}~\bibnamefont {Windey}}, \bibinfo {author} {\bibfnamefont
  {F.}~\bibnamefont {Tebbenjohanns}},\ and\ \bibinfo {author} {\bibfnamefont
  {L.}~\bibnamefont {Novotny}},\ }\bibfield  {title} {\bibinfo {title} {Ghz
  rotation of an optically trapped nanoparticle in vacuum},\ }\href
  {https://doi.org/10.1103/PhysRevLett.121.033602} {\bibfield  {journal}
  {\bibinfo  {journal} {Physical review letters}\ }\textbf {\bibinfo {volume}
  {121}},\ \bibinfo {pages} {033602} (\bibinfo {year} {2018})}\BibitemShut
  {NoStop}%
\bibitem [{\citenamefont {Chen}\ \emph {et~al.}(2019)\citenamefont {Chen},
  \citenamefont {Li},\ and\ \citenamefont {Yin}}]{chen2019nonadiabatic}%
  \BibitemOpen
  \bibfield  {author} {\bibinfo {author} {\bibfnamefont {X.-Y.}\ \bibnamefont
  {Chen}}, \bibinfo {author} {\bibfnamefont {T.}~\bibnamefont {Li}},\ and\
  \bibinfo {author} {\bibfnamefont {Z.-Q.}\ \bibnamefont {Yin}},\ }\bibfield
  {title} {\bibinfo {title} {Nonadiabatic dynamics and geometric phase of an
  ultrafast rotating electron spin},\ }\href
  {https://doi.org/10.1016/j.scib.2019.02.018} {\bibfield  {journal} {\bibinfo
  {journal} {Science Bulletin}\ }\textbf {\bibinfo {volume} {64}},\ \bibinfo
  {pages} {380} (\bibinfo {year} {2019})}\BibitemShut {NoStop}%
\bibitem [{\citenamefont {Ma}\ \emph {et~al.}(2021)\citenamefont {Ma},
  \citenamefont {Kim},\ and\ \citenamefont {Stickler}}]{ma2021torque}%
  \BibitemOpen
  \bibfield  {author} {\bibinfo {author} {\bibfnamefont {Y.}~\bibnamefont
  {Ma}}, \bibinfo {author} {\bibfnamefont {M.}~\bibnamefont {Kim}},\ and\
  \bibinfo {author} {\bibfnamefont {B.~A.}\ \bibnamefont {Stickler}},\
  }\bibfield  {title} {\bibinfo {title} {Torque-free manipulation of
  nanoparticle rotations via embedded spins},\ }\href
  {https://doi.org/10.1103/PhysRevB.104.134310} {\bibfield  {journal} {\bibinfo
   {journal} {Physical Review B}\ }\textbf {\bibinfo {volume} {104}},\ \bibinfo
  {pages} {134310} (\bibinfo {year} {2021})}\BibitemShut {NoStop}%
\bibitem [{\citenamefont {Barnett}(1915)}]{barnett1915magnetization}%
  \BibitemOpen
  \bibfield  {author} {\bibinfo {author} {\bibfnamefont {S.~J.}\ \bibnamefont
  {Barnett}},\ }\bibfield  {title} {\bibinfo {title} {Magnetization by
  rotation},\ }\href {https://doi.org/doi.org/10.1103/PhysRev.6.239} {\bibfield
   {journal} {\bibinfo  {journal} {Physical review}\ }\textbf {\bibinfo
  {volume} {6}},\ \bibinfo {pages} {239} (\bibinfo {year} {1915})}\BibitemShut
  {NoStop}%
\bibitem [{\citenamefont {Einstein}\ and\ \citenamefont
  {De~Haas}(1915)}]{einstein1915experimental}%
  \BibitemOpen
  \bibfield  {author} {\bibinfo {author} {\bibfnamefont {A.}~\bibnamefont
  {Einstein}}\ and\ \bibinfo {author} {\bibfnamefont {W.}~\bibnamefont
  {De~Haas}},\ }\bibfield  {title} {\bibinfo {title} {Experimental proof of the
  existence of amp{\`e}re’s molecular currents},\ }in\ \href@noop {} {\emph
  {\bibinfo {booktitle} {Proc. KNAW}}},\ Vol.\ \bibinfo {volume} {181}\
  (\bibinfo {year} {1915})\ p.\ \bibinfo {pages} {696}\BibitemShut {NoStop}%
\bibitem [{\citenamefont {Izumida}\ \emph {et~al.}(2022)\citenamefont
  {Izumida}, \citenamefont {Okuyama}, \citenamefont {Sato}, \citenamefont
  {Kato},\ and\ \citenamefont {Matsuo}}]{izumida2022einstein}%
  \BibitemOpen
  \bibfield  {author} {\bibinfo {author} {\bibfnamefont {W.}~\bibnamefont
  {Izumida}}, \bibinfo {author} {\bibfnamefont {R.}~\bibnamefont {Okuyama}},
  \bibinfo {author} {\bibfnamefont {K.}~\bibnamefont {Sato}}, \bibinfo {author}
  {\bibfnamefont {T.}~\bibnamefont {Kato}},\ and\ \bibinfo {author}
  {\bibfnamefont {M.}~\bibnamefont {Matsuo}},\ }\bibfield  {title} {\bibinfo
  {title} {Einstein--de haas nanorotor},\ }\href
  {https://doi.org/10.1103/PhysRevLett.128.017701} {\bibfield  {journal}
  {\bibinfo  {journal} {Physical Review Letters}\ }\textbf {\bibinfo {volume}
  {128}},\ \bibinfo {pages} {017701} (\bibinfo {year} {2022})}\BibitemShut
  {NoStop}%
\bibitem [{\citenamefont {Majorana}(1932)}]{majorana1932atomi}%
  \BibitemOpen
  \bibfield  {author} {\bibinfo {author} {\bibfnamefont {E.}~\bibnamefont
  {Majorana}},\ }\bibfield  {title} {\bibinfo {title} {Atomi orientati in campo
  magnetico variabile},\ }\href {https://doi.org/10.1007/BF02960953} {\bibfield
   {journal} {\bibinfo  {journal} {Il Nuovo Cimento (1924-1942)}\ }\textbf
  {\bibinfo {volume} {9}},\ \bibinfo {pages} {43} (\bibinfo {year}
  {1932})}\BibitemShut {NoStop}%
\bibitem [{\citenamefont {Inguscio}(2006)}]{inguscio2006pos}%
  \BibitemOpen
  \bibfield  {author} {\bibinfo {author} {\bibfnamefont {M.}~\bibnamefont
  {Inguscio}},\ }\bibfield  {title} {\bibinfo {title} {Majorana “spin-flip”
  and ultra-low temperature atomic physics},\ }\href
  {https://pdfs.semanticscholar.org/c469/3fc9a54df91d7dbc35d19bf521de39eaf467.pdf}
  {\bibfield  {journal} {\bibinfo  {journal} {PoS (EMC2006) 008}\ } (\bibinfo
  {year} {2006})}\BibitemShut {NoStop}%
\bibitem [{\citenamefont {Landau}\ and\ \citenamefont
  {Lifshitz}(1982)}]{Landau}%
  \BibitemOpen
  \bibfield  {author} {\bibinfo {author} {\bibfnamefont {L.}~\bibnamefont
  {Landau}}\ and\ \bibinfo {author} {\bibfnamefont {E.}~\bibnamefont
  {Lifshitz}},\ }\href@noop {} {\emph {\bibinfo {title} {Mechanics: Volume
  1}}}\ (\bibinfo  {publisher} {Elsevier},\ \bibinfo {year} {1982})\BibitemShut
  {NoStop}%
\bibitem [{\citenamefont {Sch{\"a}fer}\ \emph {et~al.}(2021)\citenamefont
  {Sch{\"a}fer}, \citenamefont {Rudolph}, \citenamefont {Hornberger},\ and\
  \citenamefont {Stickler}}]{schafer2021cooling}%
  \BibitemOpen
  \bibfield  {author} {\bibinfo {author} {\bibfnamefont {J.}~\bibnamefont
  {Sch{\"a}fer}}, \bibinfo {author} {\bibfnamefont {H.}~\bibnamefont
  {Rudolph}}, \bibinfo {author} {\bibfnamefont {K.}~\bibnamefont
  {Hornberger}},\ and\ \bibinfo {author} {\bibfnamefont {B.~A.}\ \bibnamefont
  {Stickler}},\ }\bibfield  {title} {\bibinfo {title} {Cooling nanorotors by
  elliptic coherent scattering},\ }\href
  {https://doi.org/10.1103/PhysRevLett.126.163603} {\bibfield  {journal}
  {\bibinfo  {journal} {Physical Review Letters}\ }\textbf {\bibinfo {volume}
  {126}},\ \bibinfo {pages} {163603} (\bibinfo {year} {2021})}\BibitemShut
  {NoStop}%
\bibitem [{\citenamefont {Ma}\ \emph {et~al.}(2020)\citenamefont {Ma},
  \citenamefont {Khosla}, \citenamefont {Stickler},\ and\ \citenamefont
  {Kim}}]{ma2020quantum}%
  \BibitemOpen
  \bibfield  {author} {\bibinfo {author} {\bibfnamefont {Y.}~\bibnamefont
  {Ma}}, \bibinfo {author} {\bibfnamefont {K.~E.}\ \bibnamefont {Khosla}},
  \bibinfo {author} {\bibfnamefont {B.~A.}\ \bibnamefont {Stickler}},\ and\
  \bibinfo {author} {\bibfnamefont {M.}~\bibnamefont {Kim}},\ }\bibfield
  {title} {\bibinfo {title} {Quantum persistent tennis racket dynamics of
  nanorotors},\ }\href {https://doi.org/10.1103/PhysRevLett.125.053604}
  {\bibfield  {journal} {\bibinfo  {journal} {Physical Review Letters}\
  }\textbf {\bibinfo {volume} {125}},\ \bibinfo {pages} {053604} (\bibinfo
  {year} {2020})}\BibitemShut {NoStop}%
\bibitem [{\citenamefont {Stickler}\ \emph {et~al.}(2021)\citenamefont
  {Stickler}, \citenamefont {Hornberger},\ and\ \citenamefont
  {Kim}}]{stickler2021quantum}%
  \BibitemOpen
  \bibfield  {author} {\bibinfo {author} {\bibfnamefont {B.~A.}\ \bibnamefont
  {Stickler}}, \bibinfo {author} {\bibfnamefont {K.}~\bibnamefont
  {Hornberger}},\ and\ \bibinfo {author} {\bibfnamefont {M.}~\bibnamefont
  {Kim}},\ }\bibfield  {title} {\bibinfo {title} {Quantum rotations of
  nanoparticles},\ }\href {https://doi.org/10.1038/s42254-021-00335-0}
  {\bibfield  {journal} {\bibinfo  {journal} {Nature Reviews Physics}\ }\textbf
  {\bibinfo {volume} {3}},\ \bibinfo {pages} {589} (\bibinfo {year}
  {2021})}\BibitemShut {NoStop}%
\bibitem [{\citenamefont {Schaff}\ \emph {et~al.}(2014)\citenamefont {Schaff},
  \citenamefont {Langen},\ and\ \citenamefont
  {Schmiedmayer}}]{schaff2014interferometry}%
  \BibitemOpen
  \bibfield  {author} {\bibinfo {author} {\bibfnamefont {J.-F.}\ \bibnamefont
  {Schaff}}, \bibinfo {author} {\bibfnamefont {T.}~\bibnamefont {Langen}},\
  and\ \bibinfo {author} {\bibfnamefont {J.}~\bibnamefont {Schmiedmayer}},\
  }\bibfield  {title} {\bibinfo {title} {Interferometry with atoms},\ }\href
  {https://doi.org/10.1393/ncr/i2014-10105-7} {\bibfield  {journal} {\bibinfo
  {journal} {La Rivista del Nuovo Cimento}\ }\textbf {\bibinfo {volume} {37}},\
  \bibinfo {pages} {509} (\bibinfo {year} {2014})}\BibitemShut {NoStop}%
\bibitem [{\citenamefont {Japha}(2021)}]{japha2021unified}%
  \BibitemOpen
  \bibfield  {author} {\bibinfo {author} {\bibfnamefont {Y.}~\bibnamefont
  {Japha}},\ }\bibfield  {title} {\bibinfo {title} {Unified model of
  matter-wave-packet evolution and application to spatial coherence of atom
  interferometers},\ }\href {https://doi.org/10.1103/PhysRevA.104.053310}
  {\bibfield  {journal} {\bibinfo  {journal} {Physical Review A}\ }\textbf
  {\bibinfo {volume} {104}},\ \bibinfo {pages} {053310} (\bibinfo {year}
  {2021})}\BibitemShut {NoStop}%
\bibitem [{\citenamefont {Barut}\ \emph {et~al.}(1992)\citenamefont {Barut},
  \citenamefont {Bo{\v{z}}i{\'c}},\ and\ \citenamefont
  {Mari{\'c}}}]{barut1992magnetic}%
  \BibitemOpen
  \bibfield  {author} {\bibinfo {author} {\bibfnamefont {A.~O.}\ \bibnamefont
  {Barut}}, \bibinfo {author} {\bibfnamefont {M.}~\bibnamefont
  {Bo{\v{z}}i{\'c}}},\ and\ \bibinfo {author} {\bibfnamefont {Z.}~\bibnamefont
  {Mari{\'c}}},\ }\bibfield  {title} {\bibinfo {title} {The magnetic top as a
  model of quantum spin},\ }\href
  {https://doi.org/10.1016/0003-4916(92)90061-P} {\bibfield  {journal}
  {\bibinfo  {journal} {Annals of Physics}\ }\textbf {\bibinfo {volume}
  {214}},\ \bibinfo {pages} {53} (\bibinfo {year} {1992})}\BibitemShut
  {NoStop}%
\bibitem [{\citenamefont {Steiner}\ \emph {et~al.}(2024)\citenamefont
  {Steiner}, \citenamefont {Pedernales},\ and\ \citenamefont
  {Plenio}}]{steiner2024pentacene}%
  \BibitemOpen
  \bibfield  {author} {\bibinfo {author} {\bibfnamefont {M.~O.}\ \bibnamefont
  {Steiner}}, \bibinfo {author} {\bibfnamefont {J.~S.}\ \bibnamefont
  {Pedernales}},\ and\ \bibinfo {author} {\bibfnamefont {M.~B.}\ \bibnamefont
  {Plenio}},\ }\bibfield  {title} {\bibinfo {title} {Pentacene-doped
  naphthalene for levitated optomechanics},\ }\href@noop {} {\  (\bibinfo
  {year} {2024})},\ \Eprint {https://arxiv.org/abs/2405.13869}
  {arXiv:2405.13869 [quant-ph]} \BibitemShut {NoStop}%
\end{thebibliography}%

\end{document}